\documentclass[12pt]{article}
\usepackage{epsfig,psfrag}

\def\nb{\bar{n}}
\def\bm#1{\mbox{\boldmath$#1$\unboldmath}}
\def\nslash{\rlap{\hspace{0.02cm}/}{n}}
\def\nbslash{\rlap{\hspace{0.02cm}/}{\bar n}}
\def\Dslash{\rlap{\hspace{0.07cm}/}{D}}

\begin{document}

\begin{titlepage}

\begin{flushright}
CLNS~06/1971\\
FERMILAB-PUB-06-242-T\\
SFB/CPP-06-32\\
SI-HEP-2006-09\\[0.2cm]
July 20, 2006
% hep-ph/0607228
\end{flushright}

\vspace{-0.3cm}
\begin{center}
\Large\bf
Factorization and Momentum-Space Resummation in Deep-Inelastic Scattering
\end{center}

\vspace{0.4cm}
\begin{center}
{\sc Thomas Becher$^a$, Matthias Neubert$^{b,c}$, and Ben D. Pecjak$^d$}\\
\vspace{0.4cm}
{\sl $^a$\,Fermi National Accelerator Laboratory\\
P.O. Box 500, Batavia, IL 60510, U.S.A.\\[0.3cm]
$^b$\,Institute for High-Energy Phenomenology\\
Newman Laboratory for Elementary-Particle Physics, Cornell University\\
Ithaca, NY 14853, U.S.A.\\[0.3cm]
$^c$\,Institut f\"ur Physik (ThEP), Johannes Gutenberg-Universit\"at\\ 
D--55099 Mainz, Germany\\[0.3cm]
$^d$\,Theoretische Physik 1, Fachbereich Physik, Universit\"at Siegen\\
D--57068 Siegen, Germany}
\end{center}

\vspace{0.2cm}
\begin{abstract}
\vspace{0.2cm}
\noindent 
Renormalization-group methods in soft-collinear effective theory are used to 
perform the resummation of large perturbative logarithms for deep-inelastic 
scattering in the threshold region $x\to 1$. The factorization theorem for the 
structure function $F_2(x,Q^2)$ for $x\to 1$ is rederived in the effective 
theory, whereby contributions from the hard scale $Q^2$ and the jet scale 
$Q^2(1-x)$ are encoded in Wilson coefficients of effective-theory operators. 
Resummation is achieved by solving the evolution equations for these 
operators. Simple analytic results for the resummed expressions are obtained 
directly in momentum space, and are free of the Landau-pole singularities 
inherent to the traditional moment-space results. We show analytically that 
the two methods are nonetheless equivalent order by order in the perturbative 
expansion, and perform a numerical comparison up to next-to-next-to-leading 
order in renormalization-group improved perturbation theory. 
\end{abstract}
\vfil

\end{titlepage}

\section{Introduction}

It is well known that fixed-order perturbation theory is not reliable for 
quantities involving several disparate scales. In such cases, higher-order 
corrections are enhanced by large logarithms of scale ratios. The standard 
solution to this problem is to split the calculation into a series of 
single-scale problems by successively integrating out the physics associated 
with the largest remaining scale. Perturbative logarithms are then resummed by 
renormalization-group (RG) evolution from the larger scales to the smaller 
ones. For collider processes, resummation is traditionally performed by other 
means, since it was not always clear how to systematically integrate out the 
physics associated with high scales in such cases. 

The simplest example of a high-energy process with a scale hierarchy which 
necessitates resummation is deep-inelastic scattering (DIS) in the threshold 
region. As the Bjorken scaling variable $x\to 1$, the invariant mass of the 
hadronic system produced in the decay, $M_X=Q\sqrt{\frac{1-x}{x}}$ (neglecting 
the nucleon mass), becomes much smaller than the momentum transfer $Q$. The 
presence of the two scales is manifest in the QCD factorization theorem 
\cite{Sterman:1986aj,Catani:1989ne,Korchemsky:1992xv}
\begin{equation}\label{eq:theorem}
   F_2^{\rm ns}(x,Q^2)
   = H(Q^2,\mu)\,Q^2\int_x^1\frac{dz}{z}\,
   J\Big( Q^2\,\frac{1-z}{z},\mu \Big)\,\frac{x}{z}\,
   \phi_q^{\rm ns}\Big( \frac{x}{z},\mu \Big) \,,
\end{equation}
for the non-singlet part of the structure function $F_2(x,Q^2)$. The result 
(\ref{eq:theorem}) is valid in the threshold region at leading power in 
$M_X^2/Q^2\approx(1-x)$ and $\Lambda_{\rm QCD}^2/{M_X^2}$. As long as 
$M_X\gg\Lambda_{\rm QCD}$, both the jet function $J(M_X^2,\mu)$ and the hard 
function $H(Q^2,\mu)$ can be evaluated in perturbation theory, whereas the 
parton distribution function $\phi_q^{\rm ns}(\xi,\mu)$ is a non-perturbative 
object. The result for the hard function involves single and (Sudakov) double 
logarithms of the form $\alpha_s^n\ln^m(Q/\mu)$, with $m\le 2n$, while the 
integral over the jet function produces logarithms $\alpha_s^n\ln^m(M_X/\mu)$. 
Irrespective of the value of the renormalization scale $\mu$, the fixed-order 
result contains large logarithms.

Traditionally, the resummation of these logarithms is performed in moment 
space. The threshold region of small $M_X$ is probed by large-$N$ moments. The 
relevant scale in Mellin space is $Q/\sqrt{N}$, so that the large perturbative 
logarithms depend on the moment parameter $N$. In 
\cite{Sterman:1986aj,Catani:1989ne} it was shown that these logarithms can be 
absorbed into a resummation exponent $G_N$, defined by integrals over two 
radiation functions $A_q(\alpha_s)$ and $B_q(\alpha_s)$,
\begin{equation}
   G_N(Q^2,\mu) = \int_0^1\!dz\,\frac{z^{N-1}-1}{1-z} \left[
   \int_{\mu^2}^{(1-z)Q^2}\!\frac{dk^2}{k^2}\,A_q(\alpha_s(k))
   + B_q\left( \alpha_s(Q\sqrt{1-z}) \right) \right] .
\end{equation}
The functions $A_q$ and $B_q$ are determined by matching with results from 
fixed-order perturbation theory and are currently known at three-loop order, 
enabling a nearly complete threshold resummation to 
next-to-next-to-next-to-leading logarithmic (N$^3$LL) accuracy 
\cite{Moch:2005ba}. The resummed momentum-space structure function 
$F_2(x,Q^2)$ is obtained from the moment-space expression by an inverse Mellin 
transformation. 

This approach to threshold resummation has several drawbacks. The first is 
related to integrations over the Landau pole in the running coupling. These 
occur twice: once in the integrals over the functions $A_q$ and $B_q$ in the 
resummation exponent, and once again when the inverse Mellin transform is 
taken to obtain results in momentum space. To perform the resummation one 
needs to specify a prescription for how to deal with these poles. Various 
methods have been proposed in the literature, such as the ``minimal 
prescription'' \cite{Catani:1996yz} or the ``tower expansion'' 
\cite{Vogt:1999xa}. The difference between these prescriptions is a 
power-suppressed effect. Since factorization theorems do receive power 
corrections, this does not appear as a problem at first sight. However, as 
discussed in \cite{Beneke:1995pq}, the Landau-pole singularity in the resummed 
expression can induce large unphysical power corrections. In the example of 
the Drell-Yan process, the ambiguity in the threshold resummation amounts to a 
power correction of order $\Lambda_{\rm QCD}/M_X$, while the physical power 
corrections to the process scale as $\Lambda_{\rm QCD}^2/M_X^2$. The fact that 
resummations with RG methods \cite{Beneke:1995pq,Korchemsky:1993uz} do not 
involve Landau-pole ambiguities illustrates that these effects do not have a 
direct physical interpretation. In particular, a Landau-pole ambiguity does 
not necessarily imply the presence of a commensurate renormalon ambiguity 
\cite{Beneke:1995pq}. Further drawbacks are that in the traditional 
resummation formalism the separation of contributions from the hard and jet 
scales is not transparent, and while the function $A_q$ has a clear 
interpretation as the cusp anomalous dimension familiar from the 
renormalization theory of Wilson lines 
\cite{Korchemsky:1987wg,Korchemskaya:1992je}, the function $B_q$ is not easily 
identified with a field-theoretical object.

In this paper we use RG techniques to perform the resummation of perturbative 
logarithms directly in momentum space. The starting point is the factorization 
formula (\ref{eq:theorem}), which we rederive using soft-collinear effective 
theory (SCET) \cite{Bauer:2000yr,Bauer:2001yt,Beneke:2002ph,Hill:2002vw}. In 
this framework, the hard function $H$ and the jet function $J$ are matching 
coefficients. The hard function arises from a first matching step, in 
which the electroweak current is matched onto a corresponding 
effective-theory current operator. In a second step, the partons associated 
with the hadronic final state are integrated out, giving rise to the jet 
function. Threshold logarithms are resummed by solving the RG equations for 
these matching functions, using techniques presented in \cite{Becher:2006nr}. 
Existing results from higher-order perturbative calculations enable us to 
perform the matching and resummation up to next-to-next-to-leading order 
(NNLO) in RG-improved perturbation theory, corresponding to the N$^3$LL 
approximation in the standard approach. We show that the results obtained in 
momentum space are formally equivalent to the more familiar moment-space 
results by deriving a formula which connects them order by order in 
perturbation theory. However, integrals over the Landau pole never appear in 
our momentum-space formulation, and the effective-theory matching functions 
and anomalous dimensions have a clear field-theoretical interpretation. 
Furthermore, we obtain a simple analytic expression for the resummed structure 
function, while the Mellin inversion which is necessary in the traditional 
approach can only be performed numerically. As a result, it is straightforward 
to match our resummed expressions onto fixed-order calculations valid outside 
the threshold region. Finally, we stress that our approach to resummation in 
$x$-space is free of the pathologies related to large unphysical power 
ambiguities found in \cite{Catani:1996yz}. In fact, it exhibits a better 
apparent perturbative convergence than the conventional approach.

In the context of SCET, the generic factorization formula for DIS has been 
discussed previously in \cite{Bauer:2002nz}, while the case $x\to 1$ has been 
studied in \cite{Manohar:2003vb,Pecjak:2005uh,Chay:2005rz,Manohar:2005az,%
Idilbi:2006dg,Chen:2006vd}. These papers make conflicting statements about the 
factorization properties of DIS in the endpoint region. Most of the 
differences are resolved after observing that, near the endpoint, the parton 
distribution function receives contributions from two distinct 
non-perturbative modes. While the two modes cannot be factorized 
perturbatively, their presence must nonetheless be taken into account to 
correctly translate the effective-theory result into the QCD factorization 
theorem (\ref{eq:theorem}). In 
\cite{Beneke:1995pq,Manohar:2003vb,Chay:2005rz,Idilbi:2006dg,Idilbi:2005ky}, 
the resummation was performed by solving the RG equations for the moments. 
This avoids Landau-pole singularities in the exponent, but as we show here, it 
is possible to solve the equations directly in momentum space. 

While threshold resummation in DIS is of limited phenomenological importance, 
it is a relatively simple process for which the perturbative results needed in 
our calculation are known at NNLO. For this reason, it provides an especially 
instructive example with which to develop our resummation formalism. However, 
many other processes fulfill factorization theorems of the same structure, in 
which the rate factorizes into a hard contribution times a jet function 
convoluted with a nonperturbative matrix element, and our formalism also 
applies to these cases. An example is heavy-particle production near 
threshold, which includes the Drell-Yan process in the limit where the 
invariant mass of the produced lepton pair is close to the center-of-mass 
energy in the collision, as well as Higgs production in the same kinematic 
region. Another example is provided by inclusive $B$-meson decays in the 
endpoint region. Our final result for the resummed DIS structure function is 
very similar to the factorized expression for radiative $B\to X_s\gamma$ decay 
as derived in \cite{Neubert:2004dd,Neubert:2005nt}. In fact, both processes 
involve the same jet function, given by the quark propagator in light-cone 
gauge. 

The outline of the paper is as follows. In Section~\ref{sec:factorization} we 
use SCET to obtain the QCD factorization formula for DIS near the endpoint,
providing a translation between the effective theory and standard discussions. 
In Section~\ref{sec:resummation} we work out the technique for threshold 
resummation in momentum space, derive a compact expression for the factorized 
structure function, and give results for the perturbative matching 
coefficients valid to NNLO in perturbation theory. In 
Section~\ref{sec:connection} we convert our results to moment space and show 
how they are connected to those obtained in the standard approach. The 
Appendix gives the perturbative expansions of the RG functions used in our 
analysis.

\section{Factorization in DIS}
\label{sec:factorization}

\begin{figure}
\begin{center}
\psfrag{a}{$$}
\psfrag{b}{$$}
\psfrag{q}{$q=P-p$}
\psfrag{P}{$p$}
\psfrag{x}{$P$}
\includegraphics[width=0.4\textwidth]{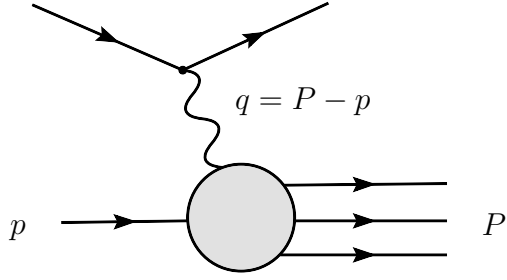} 
\end{center}
\vspace{-0.5cm}
\caption{\label{fig:kinematics}
Kinematics of DIS.}
\end{figure}

In this section we derive the QCD factorization formula for the non-singlet 
DIS structure function $F_2^{\rm ns}(x,Q^2)$, using the technology of SCET 
\cite{Bauer:2000yr,Bauer:2001yt,Beneke:2002ph,Hill:2002vw}. We consider DIS of 
electrons off a nuclear target, $e^- +N(p)\to e^- +X(P)$, as illustrated in 
Figure~\ref{fig:kinematics}. All non-trivial hadronic physics is encoded in 
the hadronic tensor
\begin{eqnarray}\label{Wdef}
   W^{\mu\nu}(p,q)
   &=& i\int d^4x\,e^{iq\cdot x}\,\langle N(p)|\,T\{ J^{\dagger\mu}(x)\,
    J^\nu(0) \}\,|N(p)\rangle \nonumber\\
   &=& \left( \frac{q^\mu q^\nu}{q^2} - g^{\mu\nu} \right) W_1
    + \left( p^\mu - q^\mu\,\frac{p\cdot q}{q^2} \right)
      \left( p^\nu - q^\nu\,\frac{p\cdot q}{q^2} \right) W_2 \,,
\end{eqnarray}
averaged over the nucleon spin. Here $J^\mu=\bar\psi\gamma^\mu\psi$ is the
electromagnetic current. The scalar functions $W_i$ can be expressed in terms 
of the kinematic invariants
\begin{equation}
   Q^2 = - q^2 \,, \qquad
   x = \frac{Q^2}{2p\cdot q} \,,
\end{equation}
where $q=P-p$ is the momentum of the virtual photon, and $x$ is the Bjorken 
scaling variable. For simplicity we focus on the flavor non-singlet component 
of the cross section, which is insensitive to the gluon distribution in the 
nucleon. It can be obtained by taking the difference of the DIS cross sections 
for scattering off different target nuclei.

%%%%%%%%%%%%%%%%%%%%%%%%%%%%%%%%%%%%%%%%%%%%%%%%%%%%%%%%%%%%%%%%%%%
%%%%%%%%%%%%%%%%%%%%%%%%%%%%%%%%%%%%%%%%%%%%%%%%%%%%%%%%%%%%%%%%%%%

The first step in analyzing the hadronic tensor is to identify which momentum 
regions give non-vanishing contributions to the Feynman diagrams for 
$W^{\mu\nu}$ in perturbative QCD. Each region is represented by a set of 
fields in SCET. The identification of regions can be done in any Lorentz 
frame. The number of modes and their relative scaling  is Lorentz 
invariant.\footnote{We disagree with the claim of \cite{Manohar:2003vb} that 
fewer momentum regions contribute in the target rest frame than in the Breit 
frame.}
Two particularly convenient reference frames are the target rest frame, where 
$p^\mu=(m,0,0,0)$ with $m$ the nucleon mass, and the Breit frame, where the 
virtual photon carries momentum $q^\mu=(0,0,0,Q)$. The two frames are related 
to each other by a Lorentz boost along the $z$-direction. Introducing the 
light-cone decomposition
\begin{equation}
   p^\mu = (n\cdot p)\,\frac{\bar n^\mu}{2}
   + (\bar n\cdot p)\,\frac{n^\mu}{2} + p_\perp^\mu
   \equiv p_+^\mu + p_-^\mu + p_\perp^\mu \,,
\end{equation}
where $n^\mu=(1,0,0,1)$ and $\bar n^\mu=(1,0,0,-1)$ are two light-like basis 
vectors ($n\cdot\bar n=2$), a generic momentum 
$p_{\rm Lab}^\mu=p_+^\mu+p_-^\mu+p_\perp^\mu$ in the target rest frame 
transforms into 
$p_{\rm Breit}^\mu=e^\eta\,p_+^\mu+e^{-\eta}\,p_-^\mu+p_\perp^\mu$ in the 
Breit frame, with the rapidity $\eta$ of the boost given by 
\begin{equation}
   e^{\pm\eta} = \frac{Q}{2mx} \left( \sqrt{1 + \frac{4m^2x^2}{Q^2}} \pm 1
   \right) .
\end{equation}
We shall discuss the different regions in the Breit frame, where the 
final-state hadronic jet moves along the $z$-direction, while the target 
nucleon moves in the opposite direction. The light-cone projections of the 
relevant momenta are (all perpendicular components vanish by choice of 
coordinates)
\begin{eqnarray}\label{scalings}
  n\cdot q &=& -Q \,, \hspace{4.38cm}
   \bar n\cdot q = Q \,, \nonumber\\
  n\cdot p &=& m\,e^\eta = \frac{Q}{x} + \frac{m^2 x}{Q} + \dots \,, \qquad
   \bar n\cdot p = m\,e^{-\eta} = \frac{m^2 x}{Q} + \dots \,, \nonumber\\
  n\cdot P &=& Q\,\frac{1-x}{x} + \frac{m^2 x}{Q} + \dots \,, \hspace{1.03cm}
   \bar n\cdot P = Q + \frac{m^2 x}{Q} + \dots \,,
\end{eqnarray}
where the neglected terms are of order $m^4 x^3/Q^3$. The invariant mass $M_X$ 
of the final-state hadronic jet is given by
\begin{equation}\label{MX2}
   M_X^2 = P^2 = Q^2\,\frac{1-x}{x} + m^2
   \approx Q^2\,\frac{1-x}{x} \gg m^2 \,.
\end{equation}
In the last step we have used that for an inclusive process the jet mass must 
be much larger than $m\sim\Lambda_{\rm QCD}$. Otherwise, the cross section 
cannot be analyzed using short-distance methods.

While the momentum of the virtual photon is fixed by kinematics, the 
final-state jet and target nucleon consist of jets of near on-shell partons, 
whose momenta have scalings consistent with the relations above. We introduce 
a small expansion parameter $\lambda\sim m/Q\sim\Lambda_{\rm QCD}/Q$ and quote 
the components $(p_+,p_-,p_\perp)$ of parton momenta in units of $Q$. Assume 
first that $x={\cal O}(1)$ is not very close to 1. Then, for the purposes of 
power counting, it follows that $q\sim Q\,(1,1,0)$, 
$p\sim Q\,(1,\lambda^2,0)$, and $P\sim Q\,(1,1,0)$. The partons making up the 
initial and final hadronic states have generic scalings
\begin{eqnarray}
   \mbox{target nucleon:} \quad
   p_{\bar c} &\sim& Q\,(1,\lambda^2,\lambda) \quad \mbox{(anti-collinear)} \,,
    \nonumber\\
   \mbox{final-state jet:} \quad
   p_h &\sim& Q\,(1,1,1) \hspace{0.65cm} \mbox{(hard)} \,,
\end{eqnarray}
where the term ``anti-collinear'' refers to collinear fields propagating in 
the negative $z$-direction. These relations change for the special case where 
$x$ is close to 1, such that $\epsilon=1-x$ becomes parametrically small. The 
momentum of the final-state jet now scales like $P\sim Q\,(\epsilon,1,0)$. 
While the valence quark in the target nucleon struck by the photon still 
carries an anti-collinear momentum 
$p_{\rm valence}\sim Q\,(1,\lambda^2,\lambda)$, the remaining partons in the 
nucleon now have total momentum scaling like 
$p-p_{\rm valence}\sim Q\,(\epsilon,\lambda^2,\lambda)$. Consequently, the 
partons making up the final-state jet and the target remnant jet have momenta 
scaling like
\begin{eqnarray}
   \mbox{final-state jet:} \quad
   p_{hc} &\sim& Q\,(\epsilon,1,\sqrt{\epsilon}) \hspace{0.85cm}
    \mbox{(hard-collinear)} \,, \nonumber\\
   \mbox{target remnants:} \quad
   p_{sc} &\sim& Q\,(\epsilon,\lambda^2,\sqrt{\epsilon}\lambda) \quad
    \mbox{(soft-collinear)} \,.
\end{eqnarray}
In these relations, the scaling of the perpendicular momentum components 
follows from the requirement that the individual partons be nearly on-shell. 
The terminology for the ``hard-collinear'' and ``soft-collinear'' modes 
follows \cite{Bosch:2003fc} and \cite{Becher:2003qh}. In the traditional 
literature on factorization in DIS 
\cite{Sterman:1986aj,Catani:1989ne,Korchemsky:1992xv} the soft-collinear modes 
were referred to as ``soft''. Relation (\ref{MX2}) implies that 
$\lambda^2\ll\epsilon\ll 1$, and there is no need to specify the relative 
scaling between $\epsilon$ and $\lambda$ in more detail. All that matters for 
the factorization analysis is that $p_{hc}^2\sim Q^2\epsilon$ is a 
perturbative scale, while $p_{\bar c}^2\sim Q^2\lambda^2$ is not. 

The discussion of factorization for the generic case, where $x={\cal O}(1)$ 
but not very close to 1, is straightforward \cite{Bauer:2002nz}. Hard modes 
are described by QCD, whereas the anti-collinear partons making up the target 
nucleon can be described in SCET. There is no need to include any other modes, 
since the only  relevant regions are hard and anti-collinear. In interactions 
of the anti-collinear fields with hard fields, only the large plus components 
$p_{\bar c +}\sim Q$ of the anti-collinear momenta should be kept at leading 
order in power counting. Correspondingly, the anti-collinear fields must be 
multipole expanded about $x_-=(\nb\cdot x) n/2$, i.e., 
$\phi_{\bar c}(x)=\phi_{\bar c}(x_-)+\dots$. Integrating out the hard modes by 
matching onto SCET yields an expression for the discontinuity of the hadronic 
tensor of the form
\begin{equation}\label{simplefact}
   \frac{1}{\pi}\,\mbox{Im}\,W^{\mu\nu}
   = \int_x^1 \frac{d\xi}{x}\,C\big(Q^2,x/\xi,\mu\big)
   \int\frac{dt}{2\pi}\,e^{-i\xi n\cdot p\,t}\,
   \langle N(p)|\,\bar\psi(tn)[tn,0]\,\gamma^\mu\,\frac{\nslash}{2}\,
   \gamma^\nu\,\psi(0)\,|N(p)\rangle \,,
\end{equation}
where $C=\delta(1-x/\xi)+{\cal O}(\alpha_s)$ is a matching coefficient in the 
effective theory, and the object $[tn,0]$ is a straight Wilson line along the 
$n$ light-cone. We have used that the SCET Lagrangian for a single collinear 
sector is equivalent to the QCD Lagrangian \cite{Bauer:2001yt} in order to 
replace the SCET fields by the usual QCD fields. The identification of the 
nucleon matrix element with the QCD parton distribution function is then 
automatic (see relation~(\ref{phidef}) below), and one arrives at the standard 
factorization formula.

The derivation of the factorization formula for $x\to 1$ is more complicated. 
It involves a two-step matching procedure similar to that used for inclusive 
semi-leptonic and radiative $B$ decays in the endpoint region 
\cite{Neubert:2004dd,Neubert:2005nt,Bosch:2003fc,Bauer:2003pi}. In a first 
matching step, hard modes are integrated out by matching QCD onto a version of 
SCET containing hard-collinear, anti-collinear, and soft-collinear fields. We 
will refer to this intermediate effective theory as SCET($hc,\bar c,sc$) for 
short. The matching function associated with this first step is the hard 
coefficient $C_V$. Because the sum of a hard-collinear momentum and an 
anti-collinear momentum has an invariant mass $(p_{hc}+p_{\bar c})^2\sim Q^2$ 
and must be counted as hard, the intermediate effective Lagrangian does not 
contain vertices coupling the hard-collinear fields to anti-collinear ones. 
These fields interact only through the exchange of soft-collinear 
``messenger'' fields. However, the soft-collinear modes can be decoupled from 
the hard-collinear ones by means of a field redefinition. After this 
decoupling, it is possible to integrate out the hard-collinear scale by 
matching onto a low-energy theory SCET($\bar c,sc$) involving only 
anti-collinear and soft-collinear modes. The matching function associated with 
this step is the jet function $J$. Having integrated out the perturbative 
modes, the final step is to evaluate the matrix element of the remaining 
operator defined in the low-energy effective theory. An important part of the 
factorization analysis is to show that this matrix element is equivalent to 
the QCD parton distribution function evaluated in the limit $x\to 1$, as 
studied e.g.\ in \cite{Sterman:1986aj,Korchemsky:1992xv}. We will show that 
this is indeed the case, and that the soft-collinear modes play an important 
role in this identification.

The appropriate Lagrangian for SCET($hc,\bar c,sc$) is a generalization of the 
effective Lagrangian for collinear and soft-collinear fields derived in 
\cite{Becher:2003qh,Becher:2003kh}. It contains hard-collinear quark and gluon 
fields $\xi_{hc}$ and $A_{hc}$, anti-collinear quark and gluon fields 
$\xi_{\bar c}$ and $A_{\bar c}$, and soft-collinear quark and gluon fields 
$\theta_{sc}$ and $A_{sc}$. The hard-collinear fields move along the 
$z$-direction, and hence $\nslash\,\xi_{hc}=0$. The anti-collinear and 
soft-collinear fields move in the opposite direction, so 
$\nbslash\,\xi_{\bar c}=0$ and $\nbslash\,\theta_{sc}=0$. The two collinear 
sectors can only interact via soft-collinear exchange, and at leading power 
only soft-collinear gluons are involved in these interactions. The 
corresponding effective Lagrangian at leading order in the expansion 
parameters $\epsilon$ and $\lambda$ is \cite{Becher:2003qh,Becher:2003kh}
\begin{eqnarray}\label{Leff}
   {\cal L}_{\rm SCET}(y)
   &=& \bar\xi_{hc}\,\frac{\nbslash}{2} \left[ in\cdot D_{hc}
    + g n\cdot A_{sc}(y_-) \right] \xi_{hc}
    - \bar\xi_{hc}\,i\Dslash_{hc\perp}\,\frac{\nbslash}{2}\,
    \frac{1}{i\bar n\cdot D_{hc}}\,i\Dslash_{hc\perp}\,\xi_{hc} \nonumber\\
   &&\mbox{}+ \bar\xi_{\bar c}\,\frac{\nslash}{2} \left[
    i\bar n\cdot D_{\bar c} + g\bar n\cdot A_{sc}(y_+) \right] \xi_{\bar c}
    - \bar\xi_{\bar c}\,i\Dslash_{\bar c\perp}\,\frac{\nslash}{2}\,
    \frac{1}{in\cdot D_{\bar c}}\,i\Dslash_{\bar c\perp}\,\xi_{\bar c}
    \nonumber\\[0.2cm]
   &&\mbox{}+ \mbox{pure glue terms + soft-collinear Lagrangian} \,,
\end{eqnarray}
where all fields without position argument are to be evaluated at the point 
$y$. The effective Lagrangian is invariant under a set of hard-collinear, 
anti-collinear, and soft-collinear gauge transformations, whose precise form 
can be found in \cite{Becher:2003qh,Beneke:2002ni}.

An important property of the SCET Lagrangian is that soft-collinear gluons can 
be decoupled from the hard-collinear and anti-collinear fields through field 
redefinitions involving Wilson lines \cite{Bauer:2001yt,Becher:2003qh}. This 
decoupling is essential for the factorization analysis below. Diagrammatic 
factorization proofs also rely on the decoupling of ``soft'' gluons from 
collinear fields. The underlying physics is that soft gluons couple to 
collinear partons through eikonal vertices, a feature explicit in the SCET 
Lagrangian (\ref{Leff}). 

\subsection{Matching of the current}

The first step in the factorization procedure is to integrate out hard 
fluctuations by matching QCD onto the intermediate effective theory 
SCET($hc,\bar c,sc$). The kinematic restrictions implied by the limit $x\to 1$ 
simplify this first matching step. Since we are dealing with the region of 
phase space where the final-state jet is hard-collinear, there are no 
contributions to the hadronic tensor where the anti-collinear partons at 
points $0$ and $x$ are connected by hard gluons. It is therefore sufficient to 
integrate out hard fluctuations at the level of the electromagnetic current. 
Time-ordered products of two currents are not needed until the second step.

We match the QCD current $J^\mu(x)=(\bar\psi\gamma^\mu\psi)(x)$ onto a current 
in SCET containing a hard-collinear quark and an anti-collinear anti-quark. 
The form of the resulting operator is dictated by gauge invariance. The 
appropriate matching relations for the QCD fields are
\begin{equation}\label{fieldmatch}
   \psi_{hc}(x)\to \big( W_{hc}^\dagger\xi_{hc} \big)(x) \,, \qquad
   \psi_{\bar c}(x)\to \big( W_{\bar c}^\dagger\xi_{\bar c} \big)(x_-) \,,
\end{equation}
where $W_{hc}$ is the hard-collinear Wilson line
\begin{equation}\label{eq:Whc}
   W_{hc}(x) = {\rm\bf P}\,\exp\left(ig\int_{-\infty}^0\!ds\,
   \nb\cdot A_{hc}(x+s\bar{n}) \right)
\end{equation}
along the $\bar n$-direction, and $W_{\bar c}$ is the analogous anti-collinear 
Wilson line along the $n$-direction. The multipole expansion in 
(\ref{fieldmatch}) requires some explanation. In the hadronic tensor 
(\ref{Wdef}), the points $0$ and $x$ are connected by a hard-collinear jet 
propagating through a cloud of soft-collinear partons. This implies that the 
position argument $x$ scales as a hard-collinear quantity, 
$x\sim (1,\epsilon^{-1},\epsilon^{-\frac12})$. It follows that not all 
components of the anti-collinear and soft-collinear momenta must be kept in 
the calculation of Feynman graphs in the effective theory. The minus and 
perpendicular components of anti-collinear and soft-collinear momenta are much 
smaller than the corresponding components of hard-collinear momenta and so 
should be expanded out. On the other hand, the large plus component 
$n\cdot p\sim Q$ of the target nucleon is canceled by the momentum component 
$n\cdot q$ of the current and turned into a momentum component of order 
$\epsilon Q$, which is of the same order as the plus component of a 
hard-collinear or soft-collinear momentum. For this reason, it would be 
incorrect to set $x_-=0$ in the argument of the soft-collinear fields entering 
the effective current operator, even though this is the correct multipole 
expansion of Lagrangian interactions between soft-collinear and anti-collinear 
fields \cite{Becher:2003qh}. Therefore, when matching the current operator 
onto SCET, one must multipole expand both the anti-collinear and 
soft-collinear fields about $x_-$. 

The two expressions in (\ref{fieldmatch}) are invariant under hard-collinear 
and anti-collinear gauge transformations, while under a soft-collinear gauge 
transformation both composite fields, $W_{hc}^\dagger\xi_{hc}$ and 
$W_{\bar c}^\dagger\xi_{\bar c}$, transform into $U_{sc}(x_-)$ times 
themselves. Thus, at tree level the gauge-invariant  matching relation for the 
current is
\begin{equation}
   \big( \bar\psi\gamma^\mu\psi \big)(x)
   \to \big( \bar\xi_{\bar c} W_{\bar c} \big)(x_-)\,\gamma_\perp^\mu
   \big( W_{hc}^\dagger\xi_{hc} \big)(x) \,.
\end{equation}
Only a single Dirac structure is possible for massless quarks. Beyond tree 
level the matching relation at leading power gets generalized to (see the 
analogous discussions in \cite{Beneke:2002ph,Becher:2003kh})
\begin{eqnarray}\label{CVdef}
   \big( \bar\psi\gamma^\mu\psi \big)(x)
   &\to& \int dt\,\widetilde C_V(t,n\cdot q,\mu)\,
    \big( \bar\xi_{\bar c} W_{\bar c} \big)(x_-)\,\gamma_\perp^\mu
    \big( W_{hc}^\dagger\xi_{hc} \big)(x +t\bar n) \nonumber\\
   &=& C_V(-n\cdot q\,\bar n\cdot\bm{P},\mu)\,
    \big( \bar\xi_{\bar c} W_{\bar c} \big)(x_-)\,\gamma_\perp^\mu
    \big( W_{hc}^\dagger\xi_{hc} \big)(x) \,.
\end{eqnarray}
In the first line we have used that $\bar n\cdot\partial$ derivatives of 
hard-collinear fields are unsuppressed in SCET power counting, allowing for 
arbitrary displacements of these fields along the $\bar n$ light-cone. In the 
second line, the object $\bm{P}$ is the hard-collinear momentum operator, and 
the Wilson coefficient $C_V$ is the Fourier transform of the position-space 
Wilson coefficient $\widetilde C_V$ appearing in the first line. In the case 
at hand, the relevant components $-n\cdot q\approx\bar n\cdot P\approx Q$ are 
fixed by kinematics (see the relations (\ref{scalings})), and so we may write 
$C_V(Q^2,\mu)$ for simplicity.

\subsection{Matching of the hadronic tensor}

The next step in the matching procedure is to evaluate the hadronic tensor in 
the intermediate effective theory. Inserting the SCET current (\ref{CVdef}) 
into (\ref{Wdef}), we find the leading-power expression
\begin{eqnarray}\label{step1}
   W^{\mu\nu}(p,q) &\to& |C_V(Q^2,\mu)|^2\,i\int d^4x\,e^{iq\cdot x} \\
   &\times& \langle N(p)|\,T\big\{
    \big( \bar\xi_{\bar c} W_{\bar c} \big)(x_-)\,\gamma_\perp^\mu
    \big( W_{hc}^\dagger\xi_{hc} \big)(x) \big( \bar\xi_{hc} W_{hc} \big)(0)\,
    \gamma_\perp^\nu \big( W_{\bar c}^\dagger \xi_{\bar c} \big)(0) \big\}\,
    |N(p)\rangle \,. \nonumber
\end{eqnarray}
The interactions of soft-collinear gluons with hard-collinear fields in 
(\ref{Leff}) can be removed by the field redefinitions 
\cite{Bauer:2001yt,Becher:2003qh}
\begin{equation}
   \xi_{hc}(x)\to S_n(x_-)\,\xi_{hc}^{(0)}(x) \,, \qquad
   A_{hc}^\mu(x)\to S_n(x_-)\,A_{hc}^{\mu(0)}(x)\,S_n^\dagger(x_-) \,,
\end{equation}
which imply $(W_{hc}^\dagger\xi_{hc})(x)\to%
S_n(x_-)\,(W_{hc}^{(0)\dagger}\xi_{hc}^{(0)})(x)$. Here
\begin{equation}\label{eq:Sn}
   S_n(x) = {\rm\bf P}\,\exp\left(ig\int_{-\infty}^0\!ds\,
   n\cdot A_{sc}(x+sn) \right)
\end{equation}
is a soft-collinear Wilson line along the $n$-direction. The redefined 
hard-collinear fields with superscripts ``(0)'' are decoupled from 
soft-collinear fields and thus interact only among themselves. After the field 
redefinition the hadronic matrix element in (\ref{step1}) factorizes into a 
vacuum matrix element of hard-collinear fields and a nucleon matrix element of 
anti-collinear and soft-collinear fields. In the second matching step, we 
``integrate out'' the hard-collinear fields, which can be done using 
perturbation theory because the hard-collinear scale is a short-distance 
scale, $p_{hc}^2\sim Q^2(1-x)\gg\Lambda_{\rm QCD}^2$. Since in a single 
(hard-)collinear sector SCET is equivalent to full QCD \cite{Beneke:2002ph}, 
the vacuum matrix element of hard-collinear fields can be rewritten in terms 
of the QCD matrix element \cite{Becher:2006qw}
\begin{eqnarray}\label{jetfun}
   \langle 0|\,T\big\{ \big( W_{hc}^{(0)\dagger}\xi_{hc}^{(0)} \big)(x)\,
    \big( \bar\xi_{hc}^{(0)} W_{hc}^{(0)} \big)(0) \big\}\,|0\rangle
   &=& \langle 0|\,T\left[ \frac{\nslash\nbslash}{4}\,W^\dagger(x)\,\psi(x)\,
    \bar\psi(0)\,W(0)\,\frac{\nbslash\nslash}{4} \right] |0\rangle
    \nonumber\\
   &=& \int\frac{d^4k}{(2\pi)^4}\,e^{-ik\cdot x}\,\frac{\nslash}{2}\,
    \bar n\cdot k\,{\cal J}(k^2,\mu) \,.
\end{eqnarray}
The object $W(x)$ denotes a Wilson line analogous to (\ref{eq:Whc}) but with 
gauge fields in full QCD. Color indices are suppressed; the correlator is 
proportional to the unit matrix in color space. We define the jet function 
through the imaginary part of ${\cal J}$ as (see e.g.\ \cite{Bosch:2004th})
\begin{equation}\label{eq:jetfun1}
   J(p^2,\mu) = \frac{1}{\pi}\,{\rm Im}\left[{i\cal J}(p^2,\mu)\right] .
\end{equation}
The jet function has support for $p^2>0$.

%%%%%%%%%%%%%%%%%%%%%%%%%%%%%%%%%%%%%%%%%%%%%%%%%%%%%%%%%%%%%%%%%%%
\begin{figure}
\begin{center}
\begin{tabular}{c}
\includegraphics[width=.95\textwidth]{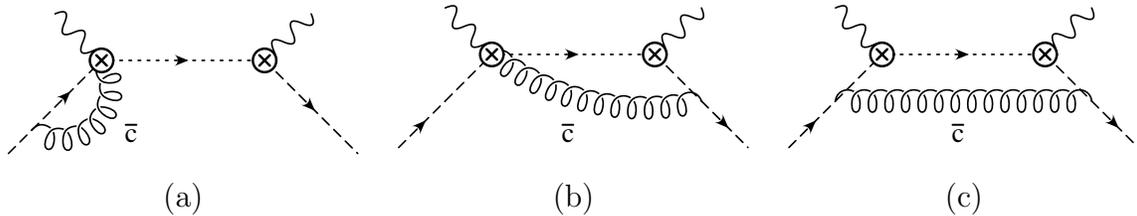}
\end{tabular}
\end{center}
\vspace{-0.8cm}
\caption{\label{fig:forbidden}
Examples of diagrams involving anti-collinear gluon exchange. The dashed 
(dotted) lines represent anti-collinear (hard-collinear) quark lines. The 
wavy lines represent the electromagnetic currents. In graph~(a) the 
anti-collinear gluon is part of the initial-state (nucleon) jet and the 
final-state propagator is hard-collinear, as required in the effective theory. 
In graphs~(b) and (c) the anti-collinear gluon is part of the final-state jet, 
whose invariant mass then becomes hard. Graphs~(b) and (c) are therefore not
part of the effective-theory representation of the hadronic tensor as 
$x\to 1$.}
\end{figure}
%%%%%%%%%%%%%%%%%%%%%%%%%%%%%%%%%%%%%%%%%%%%%%%%%%%%%%%%%%%%%%%%%%%

At this point it is important to emphasize a subtlety related to the matching 
of the forward-scattering amplitude in full QCD onto operator matrix elements 
in SCET. The anti-collinear composite fields $W_{\bar c}^\dagger\xi_{\bar c}$ 
and $\bar\xi_{\bar c} W_{\bar c}$ in (\ref{step1}) are not allowed to 
communicate via anti-collinear particle exchanges, but only through exchanges 
of soft-collinear partons. The exchange of anti-collinear particles between 
the two currents in (\ref{Wdef}) is kinematically forbidden in the region 
$x\to 1$, as this would lead to a final-state invariant hadronic mass 
$M_X\sim Q$. Since the intermediate state is hard instead of hard-collinear, 
diagrams such as those shown in Figure~\ref{fig:forbidden}(b) and (c) are not 
part of the effective-theory representation of the hadronic tensor in the 
region $x\to 1$. Such ``forbidden'' graphs are nonetheless generated (and 
yield non-vanishing results) if the SCET Feynman rules used for the matching 
of the electromagnetic current are naively applied to the hadronic tensor. We 
can construct a set of Feynman rules appropriate for the hadronic tensor by 
introducing different anti-collinear fields for the ``in'' and ``out'' states 
in the forward-scattering amplitude, and restricting interactions between the 
two anti-collinear sectors to soft-collinear exchange. These effective-theory 
Feynman rules produce graphs such as that in Figure~\ref{fig:forbidden}(a),
but not those in Figure~\ref{fig:forbidden}(b) and (c). For simplicity of 
notation, we will suppress the ``in'' and ``out'' labels on the anti-collinear 
fields, but one must make this distinction when evaluating the hadronic tensor 
in the effective theory.

After integrating out the hard-collinear fields, the resulting nucleon matrix
element in the low-energy effective theory can be reduced to
\begin{eqnarray}\label{gperp}
   &&\langle N(p)|\,\big( \bar\xi_{\bar c} W_{\bar c} \big)(x_-)\,
    S_n(x_-)\,\gamma_\perp^\mu\,\frac{\nslash}{2}\,\gamma_\perp^\nu\,
    S_n^\dagger(0)\,
    \big( W_{\bar c}^\dagger \xi_{\bar c} \big)(0)\,|N(p)\rangle
    \nonumber\\
   &=& - \langle N(p)|\,\big( \bar\xi_{\bar c} W_{\bar c} \big)(x_-)\,
    [x_-,0]_{sc}\,
    (g_\perp^{\mu\nu} - i\epsilon_\perp^{\mu\nu}\gamma_5)\,\frac{\nslash}{2}\,
    \big( W_{\bar c}^\dagger\, \xi_{\bar c} \big)(0)\,|N(p)\rangle \,,
\end{eqnarray}
where $[x_-,0]_{sc}=S_n(x_-)\,S_n^\dagger(0)$ is a straight Wilson line of 
soft-collinear gluon fields along the $n$ light-cone. In the second line we 
have defined the objects 
$g_\perp^{\mu\nu}=g^{\mu\nu}-\frac12\,(n^\mu \nb^\nu+\nb^\mu n^\nu)$
and $\epsilon_\perp^{\mu\nu}=\frac12\,\epsilon^{\mu \nu\alpha\beta}%
\nb_\alpha n_\beta$, and also used that $\nbslash\,\xi_{\bar c}=0$. The 
anti-symmetric structure vanishes after averaging over the nucleon spin. The 
appearance of the symmetric structure $g_\perp^{\mu\nu}$ implies the 
Callan-Gross relation $Q^2\,W_2=4x^2\,W_1$ at leading power and to all orders 
in perturbation theory. Hereafter, we thus focus on the structure function 
$W_1$.

Consider now the standard definition of the quark distribution function in QCD 
\cite{Collins:1981uw},
\begin{equation}\label{phidef}
   \phi_q^{\rm ns}(\xi,\mu)
   = \frac{1}{2\pi} \int_{-\infty}^\infty\!dt\,e^{-i\xi tn\cdot p}\,
   \langle N(p)|\,\bar\psi(tn)\,[tn,0]\,\frac{\rlap/n}{2}\,\psi(0)\,
   |N(p)\rangle \,,
\end{equation}
where $[tn,0]$ is a straight Wilson line of gauge fields in full QCD, and the 
superscript ``ns'' indicates the flavor non-singlet component of the 
distribution function. In the Breit frame, where the proton moves along the 
$\bar n$-direction, $\psi$ and $\bar\psi$ can be considered anti-collinear 
fields. For generic values of $\xi$ these fields carry only a portion of the 
proton's longitudinal momentum. The remaining portion $(1-\xi)\,n\cdot p$ is 
still large and can be shared between other anti-collinear partons exchanged 
between the two points $0$ and $tn$. A different picture is called for in the 
limit $\xi\to 1$, where the anti-collinear valence quarks $\psi$ and 
$\bar\psi$ carry almost all of the longitudinal momentum 
\cite{Korchemsky:1992xv}. In this case, the residual momentum component 
$(1-\xi)\,n\cdot p$ is small, and the remaining partons are soft-collinear. 
Each valence quark is described by an anti-collinear jet propagating through 
the soft-collinear cloud made up of the remaining partons. The two 
anti-collinear jets communicate through soft-collinear gluon exchange only. 

The distinct roles played by the valence and remaining partons as $\xi\to 1$ 
make it appropriate to introduce an effective field-theory description for the 
parton distribution function, in which it is matched onto an operator 
involving anti-collinear and soft-collinear fields in SCET. The most general, 
gauge-invariant form the relation (\ref{phidef}) can be matched onto in the
$\xi \to 1$ limit reads
\begin{equation}\label{phidef2}
   \phi_q^{\rm ns}(\xi,\mu) \big|_{\xi\to 1}
   = \frac{1}{2\pi} \int_{-\infty}^\infty\!dt\,e^{-i\xi tn\cdot p}\,
   \langle N(p)|\,\big( \bar\xi_{\bar c} W_{\bar c} \big)(tn)\,[tn,0]_{sc}\,
   \frac{\nslash}{2}\,\big( W_{\bar c}^\dagger \xi_{\bar c} \big)(0)\,
   |N(p)\rangle \,.
\end{equation}
It is understood that the anti-collinear fields located at the points $0$ and 
$tn$ interact only via soft-collinear gluon exchange. Both (\ref{phidef2}) and 
the QCD matrix element (\ref{phidef}) depend on the single invariant 
$p^2=m^2$, so there is no non-trivial hard matching coefficient. The matrix 
element (\ref{phidef2}) is precisely the object we encountered in 
(\ref{gperp}). We can use this correspondence along with some simple algebra 
to find 
\begin{equation}\label{W1res}
   W_1= |C_V(Q^2,\mu)|^2\,i\int d(n\cdot k)\,
   \bar n\cdot q\,{\cal J}(q^2+n\cdot k\,\bar n\cdot q,\mu)\,
   \phi_q^{\rm ns}\Big(\frac{n\cdot k}{n\cdot p},\mu \Big) \,.
\end{equation}
The structure function $F_2^{\rm ns}(x,Q^2)$ equals 
$\sum_q\,e_q^2\,x\,\frac{1}{\pi}\,\mbox{Im}\,W_1$, where the $e_q$ are quark
electric charges. Inserting the definition of the jet function 
(\ref{eq:jetfun1}), and recalling that $q^2=-Q^2$ and 
$n\cdot p\,\bar n\cdot q=Q^2/x+\mbox{power corrections}$, we obtain the final 
result for the factorization formula 
\begin{equation}\label{fact}
   F_2^{\rm ns}(x,Q^2)
   = \sum_q\,e_q^2\,|C_V(Q^2,\mu)|^2\,Q^2\,\int_x^1\!d\xi\,
   J\Big( Q^2\,\frac{\xi-x}{x},\mu \Big)\,\phi_q^{\rm ns}(\xi,\mu) \,.
\end{equation}
This formula is valid to all orders in perturbation theory and at leading 
power in $(1-x)$ and $\Lambda_{\rm QCD}^2/M_X^2$. The argument of the jet 
function takes values between 0 and $M_X^2$, where the total jet invariant 
mass was given in (\ref{MX2}). The equivalent form (\ref{eq:theorem}) is 
obtained by substituting $\xi=x/z$. At tree-level, this formula evaluates to 
the familiar parton-model expression 
$F_2^{\rm ns}(x,Q^2)=\sum_q e_q^2\,x\,\phi_q^{\rm ns}(x)$.

Relation (\ref{fact}) is the standard form of the QCD factorization formula 
for the DIS structure function in the limit $x\to 1$ 
\cite{Sterman:1986aj,Catani:1989ne,Korchemsky:1992xv}, which we have derived 
here using SCET. We hope our derivation helps resolve some of the 
disagreements in the literature. Soft-collinear messenger modes obviously play 
a crucial role in the derivation, as the parton distribution function at large 
$\xi$ is defined in terms of these fields. The proper effective-theory 
description of the parton distribution function thus requires two distinct 
non-perturbative modes. This element is missing from 
\cite{Manohar:2003vb,Manohar:2005az}, where it was argued that only one 
non-perturbative mode is needed, either because the soft graphs vanish in the 
Breit frame calculation, or because the effective-theory formulation in the 
target rest frame involves only one non-perturbative mode from the beginning. 
Although we disagree with these statements (the second of which would violate
reparameterization invariance in the effective theory), our explicit one-loop 
results agree with those derived in these papers. This is because our findings 
imply that parton evolution in the endpoint region can be described simply by 
taking the $x\to 1$ limit of the Altarelli-Parisi splitting functions, which 
is effectively what was done in the calculations of \cite{Manohar:2003vb}. Our 
explicit one-loop results also agree with those in \cite{Pecjak:2005uh}, where 
the power counting $\epsilon=1-x\sim\lambda=\Lambda_{\rm QCD}/Q$ was adopted. 
While this counting is possible and natural in view of the hierarchy 
$\lambda^2\ll \epsilon \ll 1$, it does not imply that the soft-collinear scale 
$m^2(1-x)$ depends on the scale $Q$, and the presence of this scale does not 
translate into non-perturbative $Q$-dependence in the parton distribution 
function, as was suggested in \cite{Pecjak:2005uh}. Finally, we have shown 
that the soft-collinear contributions are precisely such that they can be 
absorbed into the parton distribution function. We therefore do not confirm 
the claims of soft contributions outside the parton distribution function 
made in \cite{Chay:2005rz}. The same conclusion as ours was reached in 
\cite{Idilbi:2006dg,Chen:2006vd}, where it was argued that the infrared 
divergences due to collinear and soft emissions can be absorbed in the 
standard QCD parton distribution function, although \cite{Idilbi:2006dg} did 
not discuss how to obtain this result in the effective theory. In 
\cite{Chen:2006vd} it was claimed that there is a double-counting problem in 
SCET, which must be remedied by making certain soft subtractions from the 
collinear matrix element. We have shown here that there is no such problem. 
Double counting occurs only if one fails to notice that collinear emissions 
such as those shown in Figure~\ref{fig:forbidden}(b) and (c) must not be 
included in the effective-theory calculation near the endpoint. Similar to 
\cite{Manohar:2003vb}, the discussion in \cite{Chen:2006vd} fails to 
distinguish the virtualities of hard-collinear and anti-collinear modes, and 
it overlooks the fact that the smallest scale in the problem is not 
$Q^2(1-x)^2$ but $m^2(1-x)$. 

In the next subsection, we will emphasize the importance of soft-collinear 
Wilson loops in determining the RG properties of the effective theory.

\subsection{Decoupling transformation and cusp singularities}
\label{subsec:cusp}

%%%%%%%%%%%%%%%%%%%%%%%%%%%%%%%%%%%%%%%%%%%%%%%%%%%%%%%%%%%%%%%%%%%
\begin{figure}
\begin{center}
\psfrag{b}[b]{$t n$}
\psfrag{c}[b]{$0\phantom{a}$}
\includegraphics[width=0.3\textwidth]{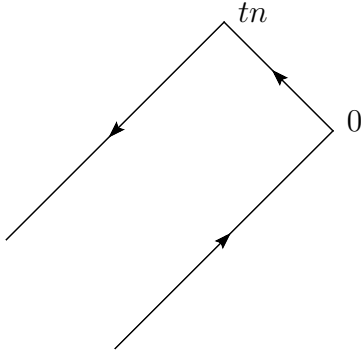}
\end{center}
\vspace{-0.5cm}
\caption{\label{wilson}
Soft-collinear Wilson line $W_C$.}
\end{figure}
%%%%%%%%%%%%%%%%%%%%%%%%%%%%%%%%%%%%%%%%%%%%%%%%%%%%%%%%%%%%%%%%%%%

An important step in the derivation of the factorization formula (\ref{fact}) 
was the identification of the parton distribution function for $\xi\to 1$ with 
the SCET matrix element on the right-hand side of (\ref{phidef2}). We can 
simplify this relation further by decoupling the soft-collinear gluons from 
the anti-collinear fields with the help of the field redefinitions
\begin{equation}
   \xi_{\bar c}(y)\to S_{\bar n}(y_+)\,\xi_{\bar c}^{(0)}(y) \,, \qquad
   A_{\bar c}^\mu(y)\to S_{\bar n}(y_+)\,A_{\bar c}^{\mu(0)}(y)\,
   S_{\bar n}^\dagger(y_+) \,,
\end{equation}
where the soft-collinear Wilson line $S_{\bar n}$ is defined in analogy with 
(\ref{eq:Sn}), but with $n$ replaced by $\bar n$. Above, $y$ is a generic 
argument of a term in the SCET Lagrangian. The multipole expansion of the 
soft-collinear fields about $y_+$ must be done everywhere except at the 
location of the current, where $x$ scales as a hard-collinear (not 
anti-collinear) quantity, see above. At this one point, we have instead
\begin{equation}\label{eq:cbardecoupling}
   \xi_{\bar c}(x_-)\to S_{\bar n}(x_-)\,\xi_{\bar c}^{(0)}(x_-) \,, \qquad
   A_{\bar c}^\mu(x_-)\to S_{\bar n}(x_-)\,A_{\bar c}^{\mu(0)}(x_-)\,
   S_{\bar n}^\dagger(x_-) \,.
\end{equation}
It follows that
\begin{equation}\label{eq27}
   \phi_q^{\rm ns}(\xi,\mu) \big|_{\xi\to 1}
   = \frac{1}{2\pi} \int_{-\infty}^\infty\!dt\,e^{-i\xi tn\cdot p}\,
   \langle N(p)|\,\big( \bar\xi_{\bar c}^{(0)} W_{\bar c}^{(0)} \big)(tn)\,
   \frac{\nslash}{2}\,W_C(t)\,
   \big( W_{\bar c}^{(0)\dagger}\xi_{\bar c}^{(0)} \big)(0)\,|N(p)\rangle \,,
\end{equation}
where
\begin{equation}\label{WC}
   W_C(t) = \langle 0|\,S_{\bar n}^\dagger(tn)\,[tn,0]_{sc}\,S_{\bar n}(0)\,
    |0\rangle
   = \langle 0|\,S_{\bar n}^\dagger(tn)\,S_n(tn)\,S_n^\dagger(0)\,
    S_{\bar n}(0)\,|0\rangle
\end{equation}
describes a closed Wilson loop consisting of the junction of a Wilson line 
extending from $-\infty$ to 0 along the $\bar n$-direction, a finite-length 
segment from 0 to $tn$ along the $n$-direction, and another Wilson line from 
$tn$ to $-\infty$ along the $\bar n$-direction, see Figure~\ref{wilson}. 
Anti-collinear virtual particles can be exchanged inside the brackets 
$(W_{\bar c}^{(0)\dagger}\xi_{\bar c}^{(0)})$ and 
$(\bar\xi_{\bar c}^{(0)} W_{\bar c}^{(0)})$ in (\ref{eq27}) but not between 
them, see Figure~\ref{fig:forbidden}. These exchanges give rise to 
non-perturbative renormalization factors $Z^{\frac12}(m,\mu)\,u(p)$, where 
$u(p)$ is an on-shell spinor, and the only invariant is $p^2=m^2$. The above 
formula then reduces to
\begin{equation}\label{phidef3}
   \phi_q^{\rm ns}(\xi,\mu) \big|_{\xi\to 1}
   = Z(m,\mu)\,\frac{n\cdot p}{2\pi} \int_{-\infty}^\infty\!dt\,
   e^{-i\xi tn\cdot p}\,W_C(t) \,.
\end{equation}
This form of the parton distribution function for $\xi\to 1$ coincides with 
eq.~(2.12) of \cite{Korchemsky:1992xv} (where our factor $Z$ is called $H$).
Performing the same decoupling transformation (\ref{eq:cbardecoupling}) on the 
SCET current (\ref{CVdef}) yields
\begin{equation}\label{Jfinal}
   \big( \bar\psi\gamma^\mu\psi \big)(x)
   \to C_V(Q^2,\mu)\,
   \big( \bar\xi_{\bar c}^{(0)} W_{\bar c}^{(0)} \big)(x_-)\,
   S_{\bar n}^\dagger(x_-)\,S_n(x_-)\,\gamma^\mu
   \big( W_{hc}^{(0)\dagger}\xi_{hc}^{(0)} \big)(x) \,.
\end{equation}
Once again, the soft-collinear fields reside in a closed Wilson loop, this 
time extending from $-\infty$ to the point $x_-$ along the $n$-direction, and 
returning to $-\infty$ along the $\bar n$-direction. 

The appearance of soft-collinear Wilson loops in the SCET representation of 
the parton distribution function (\ref{phidef3}) and the electromagnetic 
current (\ref{Jfinal}) determines their RG properties. In both cases, 
ultraviolet singularities related to Sudakov double logarithms are governed by 
the so-called cusp anomalous dimension, $\Gamma_{\rm cusp}$, which is a 
universal quantity of perturbative QCD 
\cite{Korchemsky:1992xv,Korchemsky:1987wg,Korchemskaya:1992je}. We will 
confirm this structure in our explicit calculations below.

\subsection{Power corrections}

Our focus in this paper is on the leading-order factorization formula 
(\ref{fact}), but it is important to keep in mind that this result receives 
power corrections in the small parameters $\epsilon$ and $\lambda$. A 
discussion of power corrections using SCET has been carried out for the 
closely related case of semi-leptonic inclusive $B$ decay in the endpoint 
region, where it was found that the power corrections factorize order by order 
in $1/m_b$ \cite{Lee:2004ja,Bosch:2004cb,Beneke:2004in}. It is evident that 
the same procedure applies here, so the effective theory offers a systematic 
tool to calculate power corrections for DIS near the endpoint. 

The procedure involves the two-step matching used in the leading-order case. A 
systematic treatment of the dominant power corrections requires to match the 
SCET Lagrangian and electromagnetic current at subleading order in $\epsilon$ 
and $\lambda$. We will not perform this complete matching here, but instead 
limit ourselves to a qualitative discussion of the two types of power 
corrections: those to the jet function, which can be calculated 
perturbatively, and those to the parton distribution function, which cannot. 
We give examples of each, and explain how to obtain them with effective 
field-theory techniques.

%%%%%%%%%%%%%%%%%%%%%%%%%%%%%%%%%%%%%%%%%%%%%%%%%%%%%%%%%%%%%%%%%%%
\begin{figure}
\begin{center}
\includegraphics[width=0.95\textwidth]{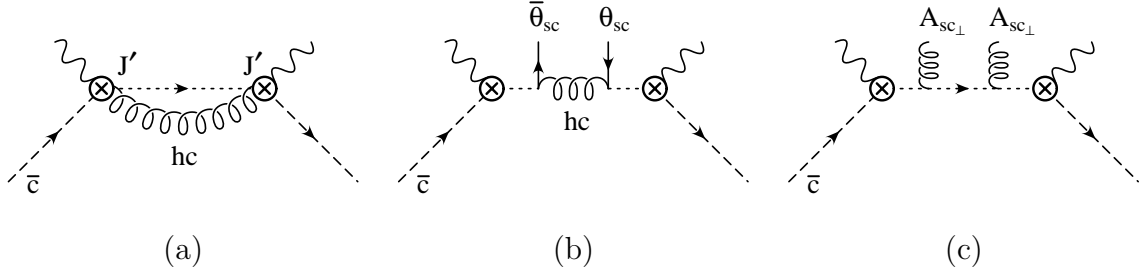}
\end{center}
\vspace{-0.8cm}
\caption{\label{fig:power}
Examples of subleading time-ordered products in SCET which give rise to power 
corrections. Graph~(a) leads to a perturbatively calculable power-suppressed 
jet function, while (b) and (c) lead to subleading parton distribution 
functions depending on three light-cone variables.}
\end{figure}
%%%%%%%%%%%%%%%%%%%%%%%%%%%%%%%%%%%%%%%%%%%%%%%%%%%%%%%%%%%%%%%%%%%

Corrections to the jet function first appear at order $(1-x)\,\alpha_s(\mu_i)$ 
and can be calculated perturbatively. To identify the full set of such 
corrections, one must match the electromagnetic current onto SCET up to order 
$\epsilon$. Time-ordered products of the subleading SCET currents containing 
extra hard-collinear fields compared to the leading-order result (\ref{CVdef}) 
build up a set of power-suppressed jet functions convoluted with the 
leading-order parton distribution. As an example, consider the time-ordered 
product $J^{\prime\dagger}(x)\,J'(0)$ involving two insertions of the 
${\cal O}(\sqrt\epsilon)$ suppressed current $J'=\nb^\mu%
\bar\xi_{\bar c} W_{\bar c} W_{hc}^\dagger i\Dslash_{hc\perp}\xi_{hc}$. The 
relevant one-loop diagram is shown in Figure~\ref{fig:power}(a). Decoupling 
the soft-collinear fields and factorizing them into the nucleon matrix element 
leaves the following vacuum matrix element of hard-collinear fields
\begin{equation}\label{subjet}
   \langle 0|\,T\big\{ \big( W_{hc}^{(0)\dagger} \Dslash_{hc\perp}^{(0)} 
    \xi_{hc}^{(0)} \big)(x)\,\big( \bar\xi_{hc}^{(0)}
    {\overleftarrow\Dslash}_{hc\perp}^{(0)} W_{hc}^{(0)} \big)(0)
   \big\}\,|0\rangle \,.
\end{equation}
The discontinuity of the Fourier-transformed matrix element defines a 
subleading jet function scaling as $(1-x)\,\alpha_s(\mu_i)$. As in the case of 
the leading-order jet function, one could equally well calculate the matrix 
element in full QCD. This particular subleading jet function has been 
discussed previously in terms of a ``non-local OPE'' very close in spirit to 
SCET in \cite{Akhoury:1998gs}, and was reconsidered in the context of SCET 
more recently in \cite{Chay:2005rz}.  

Non-perturbative power corrections are given in terms of a basis of 
subleading parton distribution functions. These are defined through nucleon 
matrix elements of power-suppressed SCET$(\bar c,sc)$ operators involving 
extra anti-collinear and soft-collinear fields compared to the leading-order 
matrix element (\ref{phidef}). To identify the complete basis, we would need 
to match both the current and the Lagrangians ${\cal L}_{\bar c+sc}$ and 
${\cal L}_{hc+sc}$ to subleading order. An important simplification can be 
made by absorbing all time-ordered products involving insertions of the 
subleading Lagrangian terms ${\cal L}_{\bar c+sc}$ into a redefinition of the 
leading-order parton distribution function $\phi(\xi)$. This can always be 
done, because such subleading parton distribution functions can only depend 
on a single light-cone variable $\xi$ and are convoluted with the same jet 
function. On a more formal level, it amounts to not treating power-suppressed 
${\cal L}_{\bar c + sc}$ terms in the interaction picture, as is normally 
done. Identifying the remaining non-perturbative structure then reduces to 
calculating power-suppressed time-ordered products involving subleading SCET 
currents or insertions of ${\cal L}_{hc+sc}$. 

The dominant subleading parton distribution function is related to the 
time-ordered product of the power-suppressed current arising from the 
multipole expansion of the anti-collinear fields, according to
$(W_{\bar c}^\dagger\xi_{\bar c})(x)=[1+x_\perp\cdot \partial_{\perp}+\frac12%
(x_\perp\cdot\partial_\perp)^2+\dots](W_{\bar c}^\dagger\xi_{\bar c})(x_-)$.
Since $x$ is a hard-collinear quantity, 
$x_\perp\cdot\partial_\perp\phi_{\bar c}%
\sim(\lambda/\sqrt{\epsilon})\,\phi_{\bar c}$. Thus the time-ordered product 
of $J^{\prime\prime\dagger}(x)\,J(0)$, with
\begin{equation}
   J^{\prime\prime}(x)
   = \frac12\,\big( \bar\xi_{\bar c} W_{\bar c} \big)(x_-)\,
   ({\overleftarrow\partial}\cdot x_\perp)^2
   \big( W_{hc}^\dagger \xi_{hc} \big)(x)
\end{equation}
results in a power correction of order $\lambda^2/\epsilon$. Because the power 
suppression comes from the current, the associated subleading parton 
distribution function depends only on a single light-cone variable, as was the 
case at leading order.

Also interesting is the new multi-local structure related to interactions of 
soft-collinear fields with the jet. As an example, consider the time-ordered 
product
\begin{equation}
   J^{(0)\dagger}(x)\,J^{(0)}(0)\,{\cal L}_{hc+sc}'(y)\, 
   {\cal L}_{hc+sc}'(z) \,,
\end{equation}
involving two insertions of the ${\cal O}(\lambda)$ suppressed Lagrangian term 
${\cal L}_{hc+sc}'\sim\bar\theta_{sc}\,A_{hc\perp}\xi_{hc}$. The relevant 
tree-level Feynman diagram is shown in Figure~\ref{fig:power}(b). Performing 
the decoupling transformation and factorizing the soft-collinear fields leads 
to a tetra-local nucleon matrix element. An additional tetra-local correction
of the same order is related to the emission of two transverse soft-collinear 
gluons, through the Lagrangian term 
${\cal L}^{\prime\prime}_{hc+sc}\sim\bar\xi_{hc}\,A_{sc\perp}\xi_{hc}$, as 
illustrated in Figure~\ref{fig:power}(c). The parton distributions 
$\phi_i(\xi_1,\xi_2,\xi_3)$ defined by the scalar decomposition of the 
Fourier-transformed matrix elements appear in a three-fold convolution with a 
perturbatively calculable jet function, to be treated as power corrections of 
order $\lambda^2\alpha_s(\mu)$. For DIS, such multi-local hadronic structures 
have not yet been considered in the literature.

\section{Renormalization-group evolution and resummation}
\label{sec:resummation}

The factorization formula for the DIS structure function derived in the 
previous section contains physics associated with different momentum scales 
factorized into a hard coefficient function $C_V$, a jet function $J$, and a
non-perturbative parton distribution function $\phi_q^{\rm ns}$. These three 
objects depend on a scale $\mu$ at which the corresponding effective-theory 
operators are renormalized. The hard matching coefficient and the jet function 
need to be calculated using perturbative QCD. These calculations can be done 
at fixed order only when the scale is chosen so as to avoid large logarithms: 
the function $C_V$ should be computed at a hard scale $\mu_h\sim Q$, while the 
jet function should be computed at an intermediate scale 
$\mu_i\sim M_X\sim Q\sqrt{1-x}$. The results of these calculations must then 
be evolved to the common scale $\mu$ in (\ref{fact}) by solving RG evolution 
equations. An advantage of the effective-theory formulation is that the RG 
equations can be solved directly in momentum space. The method has recently 
been outlined in \cite{Becher:2006nr}; in this section we give a more detailed 
derivation, filling in the technical steps.

\subsection{Evolution of the hard function}

We begin by discussing the evolution of the hard matching coefficient $C_V$ in 
(\ref{CVdef}). At leading power there is a single gauge-invariant SCET 
operator the QCD current can match onto, and hence there is no operator 
mixing. The exact evolution equation takes the form
\begin{equation}\label{gammaV}
   \frac{d}{d\ln\mu}\,C_V(Q^2,\mu)
   = \left[ \Gamma_{\rm cusp}(\alpha_s)\,\ln\frac{Q^2}{\mu^2}
   + \gamma^V(\alpha_s) \right] C_V(Q^2,\mu) \,,
\end{equation}
The appearance of the cusp logarithm and its coefficient can be explained 
starting from (\ref{Jfinal}) using arguments presented in 
\cite{Becher:2003kh}. This term in the evolution equation is associated with 
Sudakov double logarithms. The remaining term, $\gamma^V$, accounts for 
single-logarithmic evolution. 

The exact solution to the evolution equation (\ref{gammaV}) is
\begin{equation}\label{CVsol}
   C_V(Q^2,\mu) = \exp\left[ 2S(\mu_h,\mu) - a_{\gamma^V}(\mu_h,\mu) \right]
   \left( \frac{Q^2}{\mu_h^2} \right)^{-a_\Gamma(\mu_h,\mu)}\,
   C_V(Q^2,\mu_h) \,,
\end{equation}
where $\mu_h\sim Q$ is a hard matching scale at which the Wilson coefficient 
$C_V$ is calculated using fixed-order perturbation theory. The Sudakov 
exponent $S$ and the exponents $a_n$ are the solutions to the differential 
equations
\begin{equation}\label{dgl}
   \frac{d}{d\ln\mu}\,S(\nu,\mu)
   = - \Gamma_{\rm cusp}\big(\alpha_s(\mu)\big)\,\ln\frac{\mu}{\nu} \,,
    \qquad
   \frac{d}{d\ln\mu}\,a_\Gamma(\nu,\mu)
   = - \Gamma_{\rm cusp}\big(\alpha_s(\mu)\big) \,,
\end{equation}
and similarly for $a_{\gamma^V}$, subject to the initial conditions 
$S(\nu,\nu)=a_\Gamma(\nu,\nu)=a_{\gamma^V}(\nu,\nu)=0$ at $\mu=\nu$. These 
equations can be integrated by writing 
$d/d\ln\mu=\beta(\alpha_s)\,d/d\alpha_s$, where 
$\beta(\alpha_s)=d\alpha_s/d\ln\mu$ is the QCD $\beta$-function. This yields 
the exact solutions \cite{Neubert:2004dd,Bosch:2003fc}
\begin{equation}\label{RGEsols}
   S(\nu,\mu) = - \int\limits_{\alpha_s(\nu)}^{\alpha_s(\mu)}\!
    d\alpha\,\frac{\Gamma_{\rm cusp}(\alpha)}{\beta(\alpha)}
    \int\limits_{\alpha_s(\nu)}^\alpha
    \frac{d\alpha'}{\beta(\alpha')} \,, \qquad
   a_\Gamma(\nu,\mu) = - \int\limits_{\alpha_s(\nu)}^{\alpha_s(\mu)}\!
    d\alpha\,\frac{\Gamma_{\rm cusp}(\alpha)}{\beta(\alpha)} \,, 
\end{equation}
and similarly for the function $a_{\gamma^V}$. The perturbative expansions of 
the anomalous dimensions and the resulting expressions for the evolution 
functions valid at NNLO in RG-improved perturbation theory are given in the 
Appendix.

\subsection{Evolution of the jet function}

The RG evolution of the jet function is more complicated. It was recently 
shown that the exact integro-differential evolution equation obeyed by the 
function $J(p^2,\mu)$ is \cite{Becher:2006qw}
\begin{equation}\label{Jrge}
   \frac{dJ(p^2,\mu)}{d\ln\mu}
   = - \left[ 2\Gamma_{\rm cusp}(\alpha_s)\,\ln\frac{p^2}{\mu^2}
   + 2\gamma^J(\alpha_s) \right] J(p^2,\mu)
   - 2\Gamma_{\rm cusp}(\alpha_s) \int_0^{p^2}\!dp^{\prime 2}\,
   \frac{J(p^{\prime 2},\mu)-J(p^2,\mu)}{p^2-p^{\prime 2}} \,.
\end{equation} 
We encounter again the cusp anomalous dimension, and in addition a new 
function $\gamma^J$, which has been calculated at two-loop order in 
\cite{Becher:2006qw}. 

An important object in the derivation of the solution to this equation is the 
associated jet function $\widetilde j(L,\mu)$, where $L=\ln(Q^2/\mu^2)$. This 
function has originally been defined in terms of an integral over the jet 
function followed by a certain replacement rule \cite{Neubert:2005nt}. More 
elegantly, the associated jet function can be obtained from $J$ by the Laplace 
transformation
\begin{equation}\label{Laplace}
   \widetilde j\Big( \ln\frac{Q^2}{\mu^2},\mu \Big)
   = \int_0^\infty\!dp^2\,e^{-s p^2}\,J(p^2,\mu) \,, \qquad
   s = \frac{1}{e^{\gamma_E} Q^2} \,.
\end{equation}
The inverse transformation is
\begin{equation}\label{Laplaceinvert}
   J(p^2,\mu)
   = \frac{1}{2\pi i} \int_{c-i\infty}^{c+i\infty}\!ds\,e^{s p^2}\,\,
   \widetilde j\Big( \ln\frac{1}{e^{\gamma_E} s\,\mu^2},\mu \Big) \,,
\end{equation}
where the contour must be chosen to stay to the right of all discontinuities 
(i.e., $c>0$). Using the evolution equation (\ref{Jrge}) for the jet function, 
we find that the associated jet function obeys the RG equation
\begin{equation}\label{jtildeevol}
   \frac{d}{d\ln\mu}\,\widetilde j\Big( \ln\frac{Q^2}{\mu^2},\mu \Big)
   = - \left[ 2\Gamma_{\rm cusp}(\alpha_s)\,\ln\frac{Q^2}{\mu^2}
   + 2\gamma^J(\alpha_s) \right]
   \widetilde j\Big( \ln\frac{Q^2}{\mu^2},\mu \Big) \,,
\end{equation}
which is local in $Q^2$ and analogous to the evolution equation (\ref{gammaV}) 
for the hard function. The solution to this equation reads
\begin{equation}\label{jtildesol}
   \widetilde j\Big( \ln\frac{Q^2}{\mu^2},\mu \Big) 
   = \exp\left[ - 4S(\mu_i,\mu) + 2 a_{\gamma^J}(\mu_i,\mu) \right]
   \left( \frac{Q^2}{\mu_i^2} \right)^{2a_\Gamma(\mu_i,\mu)}\,
   \widetilde j\Big( \ln\frac{Q^2}{\mu_i^2},\mu_i \Big) \,,
\end{equation}
where $a_{\gamma^J}$ is defined in analogy with (\ref{dgl}). Given this 
solution one can readily derive the solution to the complicated evolution 
equation (\ref{Jrge}) for the original jet function by using the inverse 
transformation (\ref{Laplaceinvert}). The result is 
\begin{equation}\label{jsol}
   J(p^2,\mu) = \exp\left[ - 4S(\mu_i,\mu) + 2 a_{\gamma^J}(\mu_i,\mu) \right]
   \frac{e^{-\gamma_E\eta}}{\Gamma(\eta)}
   \int_0^{p^2}\!dp^{\prime 2}\,
   \frac{J(p^{\prime 2},\mu_i)}{(\mu_i^2)^{\eta} (p^2-p^{\prime 2})^{1-\eta}}
   \,,
\end{equation}
where $\eta=2a_\Gamma(\mu_i,\mu)$. This solution is valid as long as $\eta>0$, 
which implies that $\mu<\mu_i$. Equation (\ref{jsol}) is analogous to the 
solution for the evolution equation of the $B$-meson shape function found in 
\cite{Neubert:2004dd,Bosch:2004th} using a technique developed in 
\cite{Lange:2003ff}. 

Using the connection between $J$ and $\widetilde j$ implied by Laplace 
transformation, it is possible to derive an even more elegant expression for 
the jet function $J(p^2,\mu)$, which does not involve an integral and which is 
valid for both $\mu>\mu_i$ and $\mu<\mu_i$. The result relates $J$ to the 
associated jet function $\widetilde j$ evaluated at the scale $\mu_i$, where 
it can be computed using fixed-order perturbation theory. We obtain
\begin{equation}\label{sonice}
   J(p^2,\mu) = \exp\left[ - 4S(\mu_i,\mu) + 2 a_{\gamma^J}(\mu_i,\mu) \right]
   \widetilde j(\partial_\eta,\mu_i) \left[ \frac{1}{p^2}
   \left( \frac{p^2}{\mu_i^2} \right)^\eta \right]_{\!*}\,
   \frac{e^{-\gamma_E\eta}}{\Gamma(\eta)} \,,
\end{equation}
where $\partial_\eta$ denotes a derivative with respect to the quantity 
$\eta$. The star distribution is defined as \cite{Bosch:2004th,DeFazio:1999sv}
\begin{equation}\label{star}
   \int_0^{Q^2}\!dp^2\,\left[ \frac{1}{p^2}
   \left( \frac{p^2}{\mu^2} \right)^\eta \right]_{\!*}\,f(p^2)
   = \int_0^{Q^2}\!dp^2\,\frac{f(p^2)-f(0)}{p^2}
    \left( \frac{p^2}{\mu^2} \right)^\eta
    + \frac{f(0)}{\eta} \left( \frac{Q^2}{\mu^2} \right)^\eta ,
\end{equation}
where $f(p^2)$ is a smooth test function. The subtraction term involving 
$f(0)$ is required only if $\eta<0$. For small $\eta$, the above definition 
implies the expansion
\begin{equation}
   \left[ \frac{1}{p^2} \left( \frac{p^2}{\mu^2} \right)^\eta \right]_{\!*}
   = \frac{\delta(p^2)}{\eta} + \left[ \frac{1}{p^2} \right]_{\!*}
    + \eta \left[ \frac{1}{p^2}\,\ln\frac{p^2}{\mu^2} \right]_{\!*}
    + {\cal O}(\eta^2) \,.
\end{equation}
The singularity for $\eta\to 0$ is removed by the factor $1/\Gamma(\eta)$ in 
(\ref{sonice}). In the form given above, the expression for $J(p^2,\mu)$ holds 
as long as $\eta>-1$, which is sufficient for all practical purposes. For even 
smaller values of $\eta$ it would be necessary to perform further subtractions 
in (\ref{star}) by using the double-star distributions introduced in 
\cite{Lee:2005gz}.

\subsection{Matching conditions and anomalous dimensions}

To evaluate the resummed hard and jet functions at a common factorization 
scale $\mu$ requires perturbative expressions for the matching functions 
$C_V(Q^2,\mu_h)$ and $\widetilde j(L,\mu_i)$. We extract the hard coefficient 
at a scale $\mu_h\sim Q$ in the first matching step, and the associated jet 
function at a scale $\mu_i\sim Q\sqrt{1-x}$ in the second. In this way, the 
matching functions are free of large logarithms and can be reliably computed 
in fixed-order perturbation theory. We also need perturbative expressions for 
the anomalous dimensions $\Gamma_{\rm cusp}$, $\gamma^V$, and $\gamma^J$. 

The hard matching coefficient $C_V(Q^2,\mu)$ is extracted in the first 
matching step, when the vector current in full QCD is matched onto an 
effective current built out of operators in SCET. To obtain an expression for 
the Wilson coefficient one must compute, at a given order in $\alpha_s$, 
perturbative expressions for the photon vertex function in the two theories. 
The calculation is simplified greatly by performing these calculations 
on-shell, in which case all loop graphs in the effective theory are scaleless 
and hence vanish. The bare on-shell vertex function in QCD (called the 
on-shell quark form factor) has been studied extensively in the literature. 
The form factor is infrared divergent and can be regularized using dimensional 
regularization. The bare form factor at two-loop order was calculated long ago 
\cite{Kramer:1986sg,Matsuura:1987wt,Matsuura:1988sm,Gehrmann:2005pd}, and 
recently the infrared divergent contributions have even been computed at 
three-loop order \cite{Moch:2005id}. When the (vanishing) SCET graphs are 
subtracted from the QCD result, the infrared poles in $1/\epsilon$ get 
transformed into ultraviolet poles. To obtain the matching coefficient we 
introduce a renormalization factor $Z_V$, which absorbs these poles. We then 
compute
\begin{equation}
   C_V(Q^2,\mu) = \lim_{\epsilon\to 0}\,Z_V^{-1}(\epsilon,Q^2,\mu)\,
   F_{\rm bare}(\epsilon,Q^2) \,,
\end{equation}
where on the right-hand side we must also eliminate the bare coupling constant 
in favor of the renormalized coupling $\alpha_s(\mu)$. At two-loop order, we 
find (with $L=\ln(Q^2/\mu^2)$ and $\alpha_s=\alpha_s(\mu)$)
\begin{equation}
   C_V(Q^2,\mu) = 1 + \frac{C_F\alpha_s}{4\pi}
    \left( - L^2 + 3L - 8 + \frac{\pi^2}{6} \right)
    + C_F \left( \frac{\alpha_s}{4\pi} \right)^2 \left[
    C_F H_F + C_A H_A + T_F n_f H_f \right] ,
\end{equation}
where
\begin{eqnarray}
   H_F &=& \frac{L^4}{2} - 3L^3
    + \left( \frac{25}{2} - \frac{\pi^2}{6} \right) L^2
    + \left( - \frac{45}{2} - \frac{3\pi^2}{2} + 24\zeta_3 \right) L
    + \frac{255}{8} + \frac{7\pi^2}{2} - \frac{83\pi^4}{360} - 30\zeta_3 \,,
    \nonumber\\
   H_A &=& \frac{11}{9}\,L^3
    + \left( - \frac{233}{18} + \frac{\pi^2}{3} \right) L^2
    + \left( \frac{2545}{54} + \frac{11\pi^2}{9} - 26\zeta_3 \right) L 
    \nonumber\\
   &&\mbox{}- \frac{51157}{648} - \frac{337\pi^2}{108} + \frac{11\pi^4}{45}
    + \frac{313}{9}\,\zeta_3 \,, \nonumber\\
   H_f &=& - \frac49\,L^3 + \frac{38}{9}\,L^2 
    + \left( - \frac{418}{27} - \frac{4\pi^2}{9} \right) L
    + \frac{4085}{162} + \frac{23\pi^2}{27} + \frac49\,\zeta_3 \,.
\end{eqnarray}
This result agrees with the corresponding expression given in 
\cite{Idilbi:2006dg}. The anomalous dimension of the vector current in SCET is 
obtained from the coefficient $Z_V^{(1)}$ of the $1/\epsilon$ pole term via 
the relation
\begin{equation}
   \Gamma_{\rm cusp}(\alpha_s)\,\ln\frac{Q^2}{\mu^2} + \gamma^V(\alpha_s)
   = 2\alpha_s\,\frac{\partial}{\partial\alpha_s}\,Z_V^{(1)}(Q^2,\mu) \,.
\end{equation}
Using the results of \cite{Moch:2005id} the anomalous dimension can be 
extracted at three-loop order. We reproduce the well-known expression for the 
three-loop cusp anomalous dimension $\Gamma_{\rm cusp}$ \cite{Moch:2004pa}. 
For the quantity $\gamma^V$, we obtain
\begin{equation}\label{gamV}
   \gamma^V(\alpha_s) = - \frac{2\alpha_s}{\pi} 
   - (4.68 - 0.95 n_f) \left( \frac{\alpha_s}{\pi} \right)^2
   - (23.43 - 4.05 n_f + 0.029 n_f^2) 
   \left( \frac{\alpha_s}{\pi} \right)^3 + \dots \,.
\end{equation}
The exact analytic expressions for the expansion coefficients are given in the 
Appendix.

The two-loop expression for the jet function has recently been obtained in 
\cite{Becher:2006qw} starting from expression (\ref{jetfun}), by which the jet 
function is expressed in terms of a two-point vacuum correlator in full QCD. 
Using those results, the two-loop matching condition for the 
associated jet function is found to be\footnote{Our function 
$\widetilde j(L,\mu)$ should coincide with the object ${\cal M}_{N,{\rm DIS}}$ 
derived in \cite{Idilbi:2006dg} after the substitution $L\to -{\rm L}$. 
Comparing the two expressions, we disagree with the signs of the two-loop 
${\cal O}({\rm L})$ terms with color structures $C_F C_A$ and $C_F T_F n_f$, 
and with the two-loop constant terms with color structures $C_F^2$ and 
$C_F C_A$.}
\begin{equation}
   \widetilde j(L,\mu)
   = 1 + \frac{C_F\alpha_s}{4\pi} 
   \left( 2L^2 - 3L +7 - \frac{2\pi^2}{3} \right)
   + C_F \left( \frac{\alpha_s}{4\pi} \right)^2
   \left[ C_F J_F + C_A J_A + T_F n_f J_f \right] ,
\end{equation}
where
\begin{eqnarray}
   J_F &=& 2L^4 - 6L^3
    + \left( \frac{37}{2} - \frac{4\pi^2}{3} \right) L^2
    + \left( - \frac{45}{2} + 4\pi^2 - 24\zeta_3 \right) L
    + \frac{205}{8} - \frac{97\pi^2}{12} + \frac{61\pi^4}{90} - 6\zeta_3 \,,
    \nonumber\\
   J_A &=& - \frac{22}{9}\,L^3
    + \left( \frac{367}{18} - \frac{2\pi^2}{3} \right) L^2
    + \left( - \frac{3155}{54} + \frac{11\pi^2}{9} + 40\zeta_3 \right) L
    \nonumber\\
   &&\mbox{}+ \frac{53129}{648} - \frac{155\pi^2}{36} - \frac{37\pi^4}{180}
    - 18\zeta_3 \,, \nonumber\\
   J_f &=& \frac89\,L^3 - \frac{58}{9}\,L^2 
    + \left( \frac{494}{27} - \frac{4\pi^2}{9} \right) L
    - \frac{4057}{162} + \frac{13\pi^2}{9} \,.
\end{eqnarray}
The anomalous dimension kernel entering (\ref{Jrge}) has been calculated at 
two-loop order \cite{Becher:2006qw}. In Section~\ref{subsec:Altarelli} below, 
we will show that the difference $(\gamma^J-\gamma^V)$ multiplies the 
$\delta(1-z)$ term in the Altarelli-Parisi splitting function 
$P_{q\leftarrow q}(z)$, which has recently been calculated at three-loop order 
\cite{Moch:2004pa}. When combined with (\ref{gamV}) this relation serves as a 
cross-check of the two-loop result obtained from the direct calculation in 
\cite{Becher:2006qw}, and further it can be employed to extract the three-loop 
coefficient of the jet-function anomalous dimension. We find 
\begin{equation}
   \gamma^J(\alpha_s) = - \frac{\alpha_s}{\pi} 
   - (0.364 - 0.556 n_f) \left( \frac{\alpha_s}{\pi} \right)^2
   - (3.18 - 1.33 n_f + 0.011 n_f^2) 
   \left( \frac{\alpha_s}{\pi} \right)^3 + \dots \,.
\end{equation}
The exact analytic expressions for the expansion coefficients are given in the 
Appendix.

\subsection{Resummation of large logarithms}

We are now ready to write down a resummed expression for the structure 
function $F_2^{\rm ns}(x,Q^2)$, valid to all orders in perturbation theory and 
at leading power in $(1-x)$ and $\Lambda_{\rm QCD}^2/M_X^2$. The result is
\begin{eqnarray}\label{Fres}
   F_2^{\rm ns}(x,Q^2)
   &=& \sum_q e_q^2\,|C_V(Q^2,\mu_h)|^2
    \left( \frac{Q^2}{\mu_h^2} \right)^{-2a_\Gamma(\mu_h,\mu_f)} \nonumber\\
   &&\times \exp\left[ 4S(\mu_h,\mu_f) - 4S(\mu_i,\mu_f)
    - 2a_{\gamma^V}(\mu_h,\mu_f) + 2a_{\gamma^J}(\mu_i,\mu_f) \right]
    \nonumber\\
   &&\times \widetilde j(\partial_\eta,\mu_i)\,
    \frac{e^{-\gamma_E\eta}}{\Gamma(\eta)}\,Q^2
    \int_x^1\!d\xi \left[ \frac{1}{Q^2(\xi/x-1)}
    \left( \frac{Q^2(\xi/x-1)}{\mu_i^2} \right)^\eta \right]_{\!*}\,
    \phi_q^{\rm ns}(\xi,\mu_f) \,,
\end{eqnarray}
where $\eta=2a_\Gamma(\mu_i,\mu_f)$, as above. To leading power, we could 
approximate $(\xi/x-1)\to(\xi-x)$, but we prefer to keep the full $x$ 
dependence in our numerical studies below. The ``factorization scale'' 
$\mu_f\equiv\mu$ is, by definition, the scale at which the parton distribution 
function is renormalized.

The Sudakov exponent can be simplified using the general relations
\begin{eqnarray}
   a_\Gamma(\mu_1,\mu_2) + a_\Gamma(\mu_2,\mu_3)
   &=& a_\Gamma(\mu_1,\mu_3) \,, \nonumber\\
   S(\mu_1,\mu_2) + S(\mu_2,\mu_3)
   &=& S(\mu_1,\mu_3) + \ln\frac{\mu_1}{\mu_2}\,a_\Gamma(\mu_2,\mu_3) \,.
\end{eqnarray}
Introducing the short-hand notations
\begin{equation}\label{gammaphi}
   \gamma^\phi = \gamma^J - \gamma^V \,, \qquad
   a_{\gamma^\phi} = a_{\gamma^J} - a_{\gamma^V} \,, 
\end{equation}
we find after a straightforward calculation
\begin{eqnarray}\label{generalbeauty}
   F_2^{\rm ns}(x,Q^2)
   &=& \sum_q e_q^2\,|C_V(Q^2,\mu_h)|^2
    \left( \frac{Q^2}{\mu_h^2} \right)^{-2a_\Gamma(\mu_h,\mu_i)}
    \exp\left[ 4S(\mu_h,\mu_i) - 2a_{\gamma^V}(\mu_h,\mu_i) \right] 
    \nonumber\\
   &&\times \exp\left[ 2a_{\gamma^\phi}(\mu_i,\mu_f) \right]
    \widetilde j\Big( \ln\frac{Q^2}{\mu_i^2}+\partial_\eta,\mu_i \Big)\,
    \frac{e^{-\gamma_E\eta}}{\Gamma(\eta)}\,\int_x^1\!d\xi\,
    \frac{\phi_q^{\rm ns}(\xi,\mu_f)}{\left[ \left( \xi/x-1 \right)^{1-\eta}
          \right]_{\!*}} \,.
\end{eqnarray}

The remaining integral can be performed by noting that, on very general 
grounds, the behavior of the parton distribution function near the endpoint 
can be parameterized as
\begin{equation}\label{asymp}
   \phi_q^{\rm ns}(\xi,\mu_f) \big|_{\xi\to 1}
   = {\cal N}(\mu_f)\,(1-\xi)^{b(\mu_f)} \Big[ 1 + {\cal O}(1-\xi) \Big] \,,
\end{equation}
where $b(\mu_f)>0$. We will see in the next subsection that this functional 
form is preserved under RG evolution. Defining a $K$ factor by the ratio
\begin{equation}
   K(x,Q^2,\mu_f) 
   = \frac{F_2^{\rm ns}(x,Q^2)}{\sum_q e_q^2\,x\,\phi_q^{\rm ns}(x,\mu_f)} \,, 
\end{equation}
we now obtain our final expression
\begin{eqnarray}\label{beauty}
   K(x,Q^2,\mu_f)
   &=& |C_V(Q^2,\mu_h)|^2
    \left( \frac{Q^2}{\mu_h^2} \right)^{-2a_\Gamma(\mu_h,\mu_i)}
    \exp\left[ 4S(\mu_h,\mu_i) - 2a_{\gamma^V}(\mu_h,\mu_i) \right] \\
   &\times& \exp\left[ 2a_{\gamma^\phi}(\mu_i,\mu_f) \right]
    \left(\frac{1-x}{x}\right)^\eta\,
    \widetilde j\Big( \ln\frac{Q^2}{\mu_i^2}\,\frac{1-x}{x} + \partial_\eta,
                      \mu_i \Big)\,
    \frac{e^{-\gamma_E\eta}\,\Gamma(1+b(\mu_f))}{\Gamma(1+b(\mu_f)+\eta)} \,,
    \nonumber
\end{eqnarray}
where as before $\eta=2a_\Gamma(\mu_i,\mu_f)$. In this expression, the 
dependence on the two physical scales $Q^2$ and 
$M_X^2=Q^2\,\frac{1-x}{x}$ (neglecting the nucleon mass) is completely 
explicit. Our exact result is independent of the scales $\mu_h$ and $\mu_i$, 
at which QCD is matched onto the intermediate and final effective theories, 
SCET($hc,\bar c,sc$) and SCET($\bar c,sc$), respectively. In practice, a 
residual scale dependence remains once we truncate the perturbative expansions 
of the various objects in the factorization formula. The final answer 
simplifies further if we choose the ``natural'' values of the two matching 
scales, $\mu_h=Q$ and $\mu_i=M_X$. However, we prefer to vary the matching 
scales over some reasonable range and take the variation of the results as an 
estimate of higher-order perturbative effects. Note that by definition the $K$ 
factor {\em does\/} depend on the choice of the factorization scale $\mu_f$. 
It is conventional in the literature on DIS to identify $\mu_f$ with the hard 
scale $Q$. However, from the point of view of an effective field theory it 
would be more natural to choose $\mu_f$ below the intermediate matching scale 
$\mu_i\sim Q\sqrt{1-x}$. A typical choice would be $\mu_f$ of order a few 
GeV, independent of the dynamical variables $x$ and $Q$.

\begin{table}
\centerline{\parbox{14cm}{\caption{\label{tab:counting}
Different approximation schemes for the evaluation of the resummed 
factorization formula (\ref{beauty})}}}
\vspace{0.1cm}
\begin{center}
\begin{tabular}{ccc|ccc}
\hline\hline
RG-impr.\ PT & Log.\ Approx.\ & Accuracy $\sim\alpha_s^n L^k$
 & $\Gamma_{\rm cusp}$ & $\gamma^V$, $\gamma^J$ & $C_V$, $\widetilde j$ \\
\hline
--- & LL & $n+1\le k\le 2n$~($\alpha_s^{-1}$) & 1-loop & tree-level
 & tree-level \\
LO & NLL & $n\le k\le 2n$ \hspace{0.76cm} ($\alpha_s^0$) & 2-loop & 1-loop
 & tree-level \\
NLO & NNLL & $n-1\le k\le 2n$~~($\alpha_s$) & 3-loop & 2-loop & 1-loop \\
NNLO & NNNLL & $n-2\le k\le 2n$~~($\alpha_s^2$) & 4-loop & 3-loop & 2-loop \\
\hline\hline
\end{tabular}
\end{center}
\end{table}

In (\ref{generalbeauty}) and (\ref{beauty}) we have accomplished the complete 
resummation of threshold logarithms for $F_2$ directly in momentum space. That 
the final answer is a convolution (rather than a product) of a hard-scattering 
kernel with the parton distribution function is reflected in the non-trivial 
dependence on the hadronic parameter $b(\mu_f)$ describing the large-$\xi$ 
behavior of $\phi_q^{\rm ns}(\xi,\mu_f)$. Our factorized expression for the 
DIS structure function is very similar to that for the $B\to X_s\gamma$ decay 
rate derived in \cite{Neubert:2005nt}. Although the hard functions and soft 
matrix elements differ, the jet function is the same in both cases. An 
important advantage of our momentum-space approach is that in the limit where 
the two matching scales are set equal to the factorization scale, 
$\mu_h=\mu_i=\mu_f$, the resummed results (\ref{generalbeauty}) and 
(\ref{beauty}) automatically reduce to the corresponding expressions valid in 
fixed-order perturbation theory (expanded about $x=1$). Consequently, it is 
straightforward to match our resummed expressions onto fixed-order 
calculations valid outside the threshold region. 

The right-hand sides of (\ref{generalbeauty}) and (\ref{beauty}) can be 
evaluated at any desired order in resummed perturbation theory. 
Table~\ref{tab:counting} shows what is required to obtain different levels of 
accuracy in the perturbative evaluation of the result. In this work we adopt 
the counting scheme of RG-improved perturbation theory, where at leading-order 
(LO) one includes all terms of order~1, at next-to-leading order (NLO) 
one includes all terms of order $\alpha_s$, and at next-to-next-to-leading 
order (NNLO) one includes all terms of order $\alpha_s^2$. We count the large 
logarithms $L\in\{\ln\mu_h/\mu_i,\,\ln\mu_i/\mu_f,\,\ln(1-x)\}$ like 
$1/\alpha_s$. In the literature on DIS, the LO approximation is also referred 
to as the next-to-leading logarithmic (NLL) approximation, the NLO result is 
referred to as the next-to-next-to-leading logarithmic (NNLL) approximation, 
etc. The leading logarithmic (LL) approximation is listed only for 
completeness, as it neglects terms that are parametrically much larger than 1.

%%%%%%%%%%%%%%%%%%%%%%%%%%%%%%%%%%%%%%%%%%%%%%%%%%%%%%%%%%%%%%%%%%%
\begin{figure}
\begin{center}
\begin{tabular}{lr}
\hspace{-0.05\textwidth}
\psfrag{x}[B]{$x$}
\psfrag{y}[]{}
\psfrag{Q}[]{$Q=5\,$GeV}
\psfrag{T}[]{$K(x,Q^2,\mu_f)$}
\psfrag{n}[]{$M_X<1$\,GeV}
\includegraphics[width=0.48\textwidth]{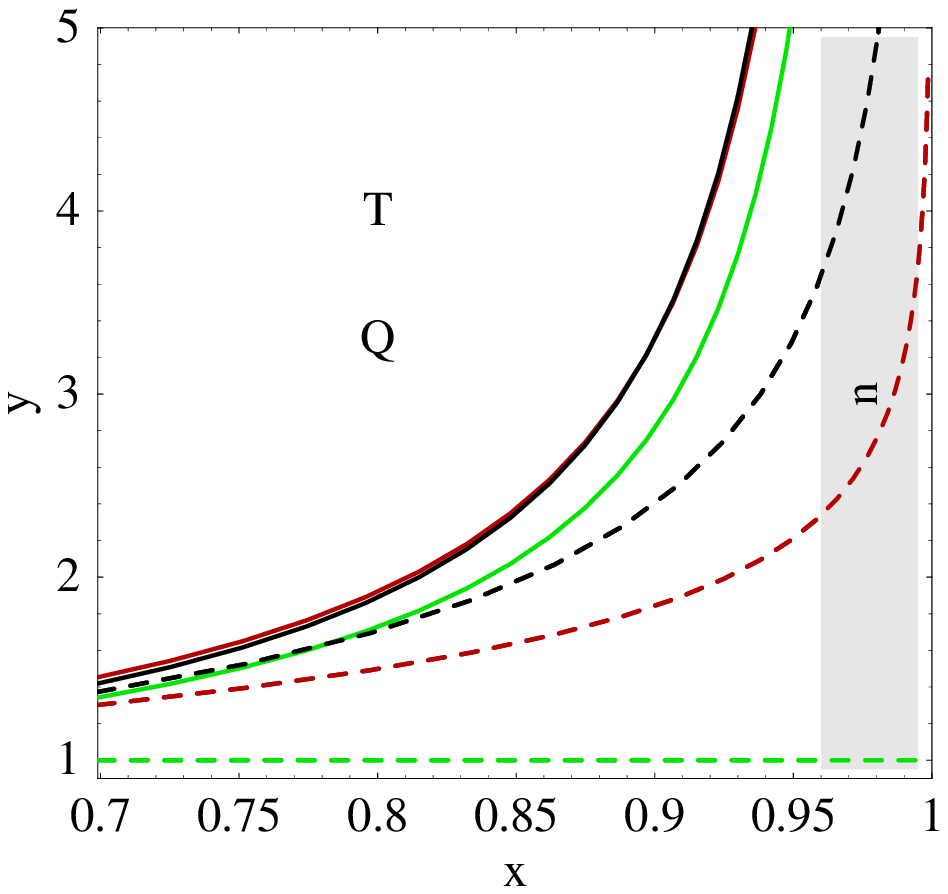} & 
\psfrag{x}[B]{$x$}
\psfrag{y}[]{}
\psfrag{Q}[]{$Q=30\,$GeV}
\psfrag{T}[]{$K(x,Q^2,\mu_f)$}
\includegraphics[width=0.48\textwidth]{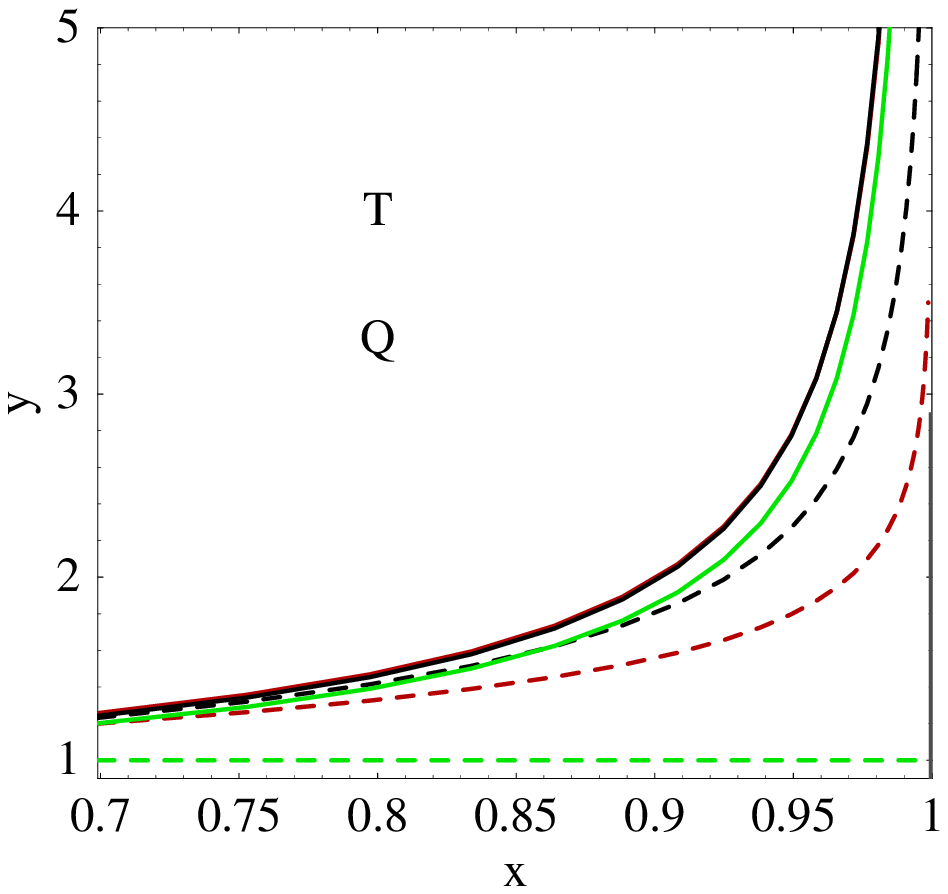}
\end{tabular}
\end{center}
\vspace{-0.5cm}
\caption{\label{comparisonFixed}
Comparison between fixed-order (dashed) and resummed results (solid) for the 
$K$ factor. The green curves are the LO result, red NLO, black NNLO. For the 
resummed result, we set $\mu_h=Q$, $\mu_i=M_X$, $\mu_f=Q$, and $b(\mu_f)=4$. 
The fixed-order result is obtained by setting all scales equal to $\mu_f$.}
\end{figure}
%%%%%%%%%%%%%%%%%%%%%%%%%%%%%%%%%%%%%%%%%%%%%%%%%%%%%%%%%%%%%%%%%%%%%

%%%%%%%%%%%%%%%%%%%%%%%%%%%%%%%%%%%%%%%%%%%%%%%%%%%%%%%%%%%%%%%%%
\begin{figure}
\begin{center}
\begin{tabular}{lr}
\hspace{-0.05\textwidth}
\psfrag{x}[B]{$x$}
\psfrag{y}[]{}
\psfrag{Q}[l]{$\quad Q=5$\,GeV}
\psfrag{T}[lt]{$\quad K(x,Q^2,\mu_f)$}
\psfrag{n}[]{$M_X<1$\,GeV}
\includegraphics[width=0.48\textwidth]{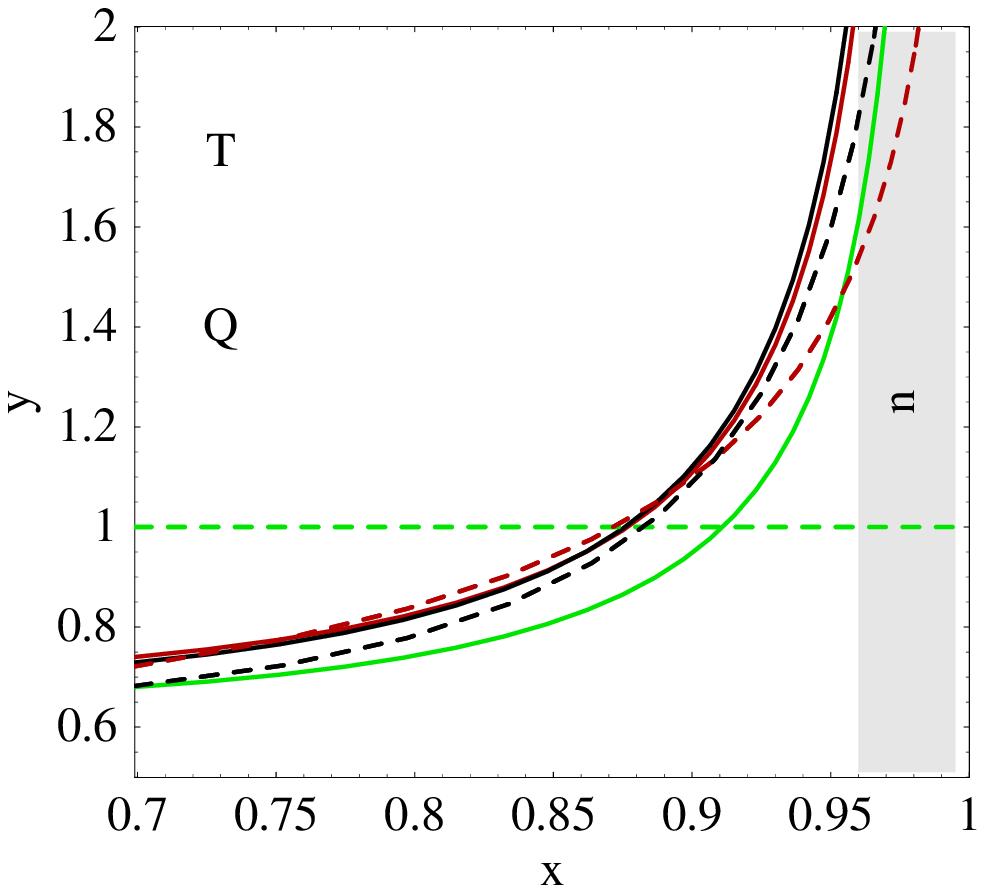} & 
\psfrag{x}[B]{$x$}
\psfrag{y}[]{}
\psfrag{Q}[l]{$\quad Q=30$\,GeV}
\psfrag{T}[lt]{$\quad K(x,Q^2,\mu_f)$}
\includegraphics[width=0.48\textwidth]{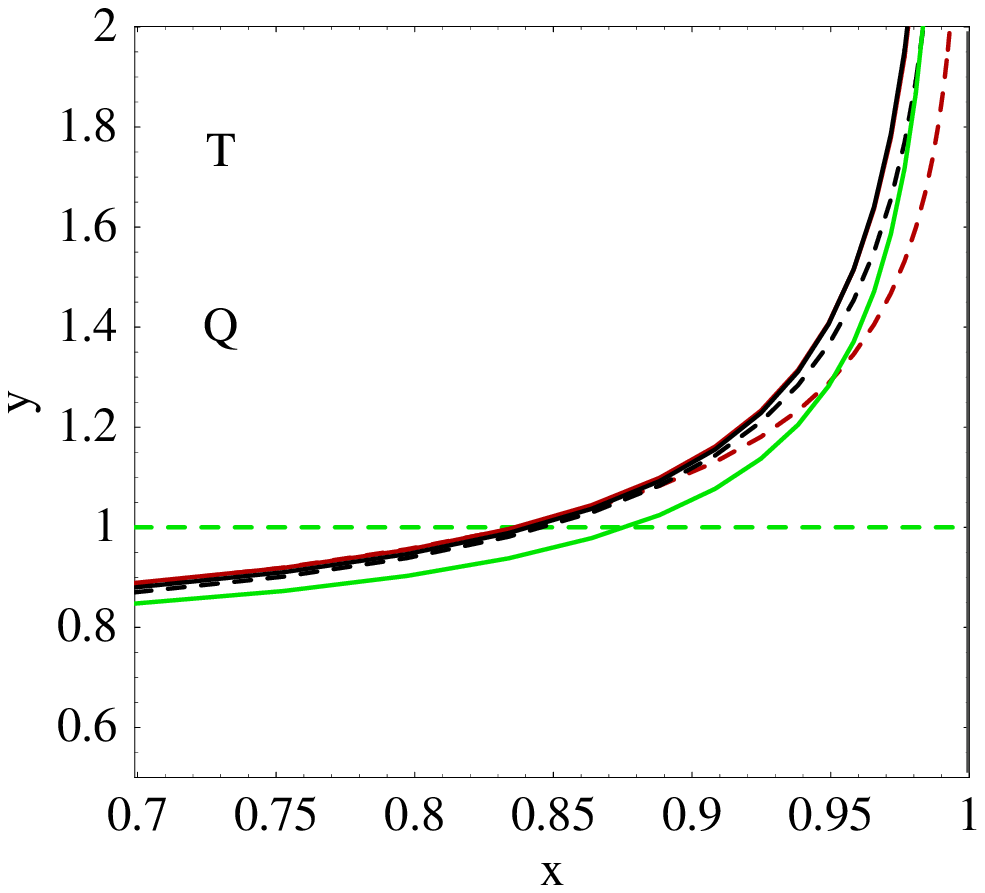}
\end{tabular}
\end{center}
\vspace{-0.5cm}
\caption{\label{comparisonLowMuf}
Same as Figure~\ref{comparisonFixed}, but with a lower choice of the 
factorization scale. Specifically, we take $\mu_f=1.5$\,GeV for $Q=5$\,GeV 
(left), and $\mu_f=10$\,GeV for $Q=30$\,GeV (right).} 
\end{figure}
%%%%%%%%%%%%%%%%%%%%%%%%%%%%%%%%%%%%%%%%%%%%%%%%%%%%%%%%%%%%%%%%%%%

In Figure~\ref{comparisonFixed}, we compare the fixed-order calculation of the 
$K$ factor with the resummed result for $Q=5$\,GeV and $Q=30$\,GeV. For the 
resummed result we use the default choice of scales $\mu_h=Q$, 
$\mu_i=M_X=Q\sqrt{\frac{1-x}{x}}$ and take the asymptotic form of the parton 
distribution (\ref{asymp}) with $b(\mu_f)=4$ in both cases. Following common 
practice we choose $\mu_f=Q$ for the factorization scale. In this case the 
quantity $\eta<0$, and because of the factor $(\frac{1-x}{x})^\eta$ in 
(\ref{beauty}) the resummed results diverge as $x$ approaches~1. The figure 
illustrates that higher-order corrections become important as $x\to 1$, and 
that fixed-order perturbation theory is no longer adequate in this limit. The 
magnitude of the $K$ factor can be reduced by adopting a lower choice for the 
factorization scale, which is more in line with the philosophy of an effective 
field-theory approach. For example, we may consider taking 
$\mu_f\approx M_X(x=0.9)\approx Q/3$, corresponding to a typical hadronic 
invariant mass in the endpoint region. The corresponding results are shown in 
Figure~\ref{comparisonLowMuf}. We observe that with such a choice of the 
factorization scale the $K$ factor takes more moderate values, and also that 
the effects of resummation are less significant.

%%%%%%%%%%%%%%%%%%%%%%%%%%%%%%%%%%%%%%%%%%%%%%%%%%%%%%%%%%%%%%%%%%%
\begin{figure}
\begin{center}
\psfrag{x}[B]{$x$}
\psfrag{y}[]{}\psfrag{0.75}[]{}\psfrag{0.85}[]{}\psfrag{0.95}[]{}
\psfrag{1.75}[]{}\psfrag{1.25}[]{}
\psfrag{a}{\phantom{ab}LO}
\psfrag{b}{\phantom{ab}NLO}
\psfrag{c}{\phantom{ab}NNLO}
\psfrag{T}{$K(x,Q^2,\mu_f)$}
\includegraphics[width=\textwidth]{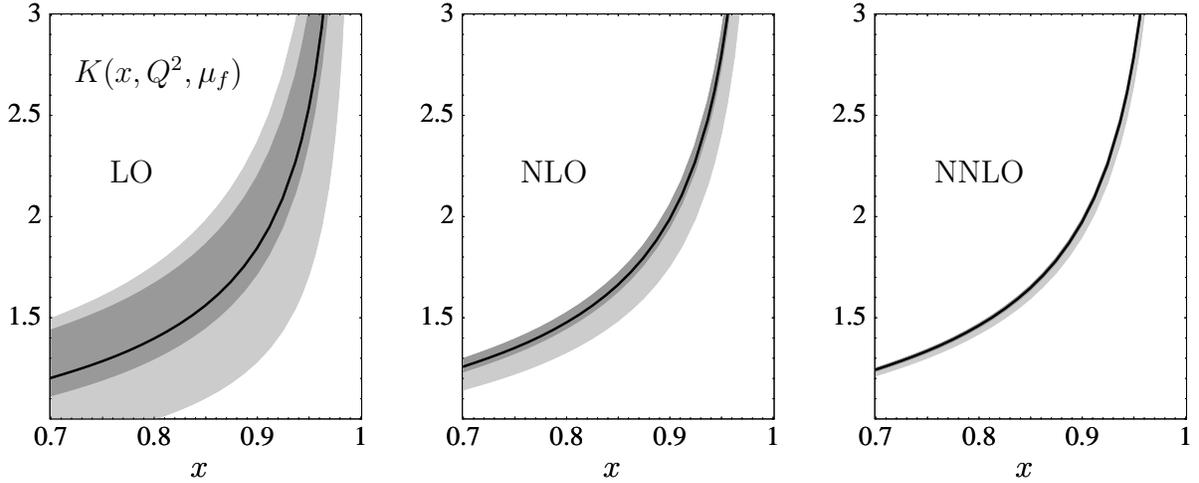} 
\end{center}
\vspace{-0.5cm}
\caption{\label{scalevar}
Scale variation of the $K$ factor at $Q=30$\,GeV. The light-gray band is 
obtained by varying $M_X/2<\mu_i<2M_X$, while the dark-gray band arises from 
varying the hard scale $Q/2<\mu_h<2Q$. We set $\mu_f=30$\,GeV and 
$b(\mu_f)=4$.}
\end{figure}
%%%%%%%%%%%%%%%%%%%%%%%%%%%%%%%%%%%%%%%%%%%%%%%%%%%%%%%%%%%%%%%%%%%

In Figure~\ref{scalevar}, we show the scale dependence of the result obtained 
by varying the hard and intermediate scales by a factor of 2 about their 
default values. The figure shows a dramatic reduction in scale uncertainty 
when going from LO to NNLO. It also suggests that varying the two matching 
scales individually by a factor of 2 may overestimate the perturbative 
uncertainty, because the higher-order results lie near the center of the large 
band obtained by varying the renormalization scales in the low-order ones. A 
variation of the scales by a factor of $\sqrt{2}$ better represents the 
uncertainty in the present case. Furthermore, it seems reasonable to perform 
the scale variations in such a way that the hierarchy of scales $\mu_h>\mu_i$ 
is preserved.

We stress that the resummation of large logarithms accomplished in 
(\ref{beauty}) is under perturbative control as long as 
$(1-x)\gg\Lambda_{\rm QCD}^2/Q^2$, since only then is the intermediate 
matching scale $\mu_i\sim Q\sqrt{1-x}\gg\Lambda_{\rm QCD}$ a short-distance 
scale. Physically, this condition is equivalent to saying that the final-state 
jet can be treated in an inclusive way using a partonic language. Numerically, 
we can assume that perturbation theory at the jet scale breaks down in the 
region where $M_X<1$\,GeV. We illustrate this boundary in our $x$-space 
results with the gray band in Figure~\ref{comparisonFixed}. For $Q=5$\,GeV, 
our approach is valid as long as $x<0.96$. For $Q=30$\,GeV, it extends all the 
way to $x<0.999$, so that in this case the band is not visible on the scale of 
the plot. While our theoretical description breaks down very close to the 
endpoint, we note that weighted integrals of the jet function over an interval 
$x\in[x_0,1]$ can be calculated starting from (\ref{generalbeauty}) as long as 
$x_0$ is in the short-distance domain.

%%%%%%%%%%%%%%%%%%%%%%%%%%%%%%%%%%%%%%%%%%%%%%%%%%%%%%%%%%%%%%%%%%%
\begin{figure}
\begin{center}
\begin{tabular}{lr}
\hspace{-0.05\textwidth}
\psfrag{x}[B]{$x$}
\psfrag{y}[]{}
\psfrag{Q}[lb]{$Q=5$\,GeV}
\psfrag{T}[rl]{\hspace{-0.5cm}$10^3\,(1-x)^4\,K(x,Q^2,\mu_f)$}
\psfrag{n}[]{$M_X<1$\,GeV}
\includegraphics[width=0.48\textwidth]{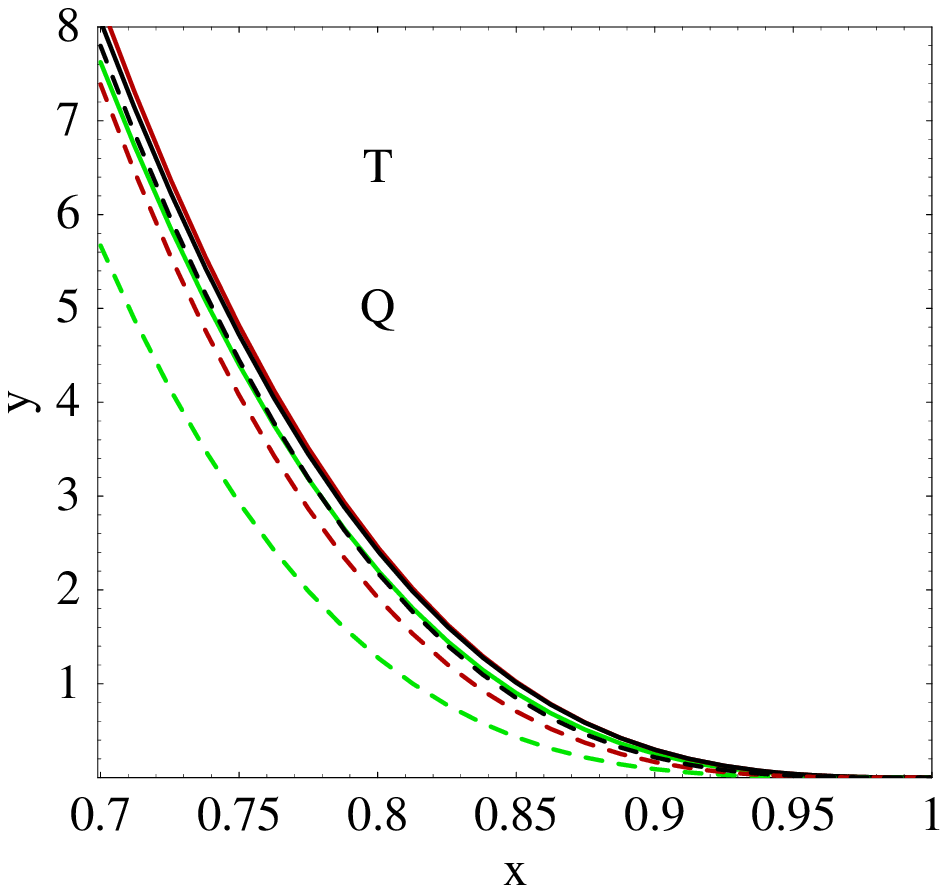} & 
\psfrag{x}[B]{$x$}
\psfrag{y}[]{}
\psfrag{Q}[lb]{$Q=30\,$GeV}
\psfrag{T}[rl]{\hspace{-0.5cm}$10^3\,(1-x)^4\,K(x,Q^2,\mu_f)$}
\includegraphics[width=0.48\textwidth]{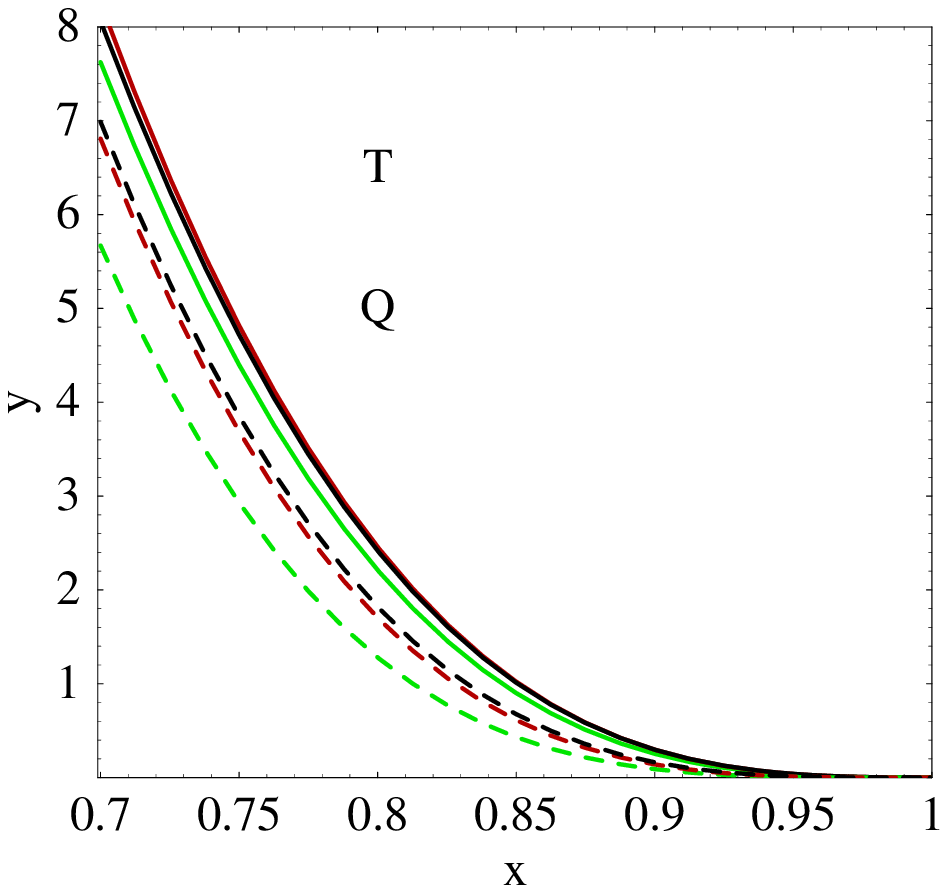}
\end{tabular}
\vspace{-0.5cm}
\end{center}
\caption{\label{comparisonFixedN}
Same as Figure~\ref{comparisonFixed}, but now including the suppression factor 
$(1-x)^b$ with $b(\mu_f)=4$ from the parton distribution function.}
\end{figure}
%%%%%%%%%%%%%%%%%%%%%%%%%%%%%%%%%%%%%%%%%%%%%%%%%%%%%%%%%%%%%%%%%%%

Experimentally, structure functions at large $x$ are very difficult to 
measure, because of the rapid decrease of the parton distribution function as 
$x\to 1$. This is illustrated in Figure~\ref{comparisonFixedN}, where we 
include the suppression factor from the parton distribution. Because of this 
strong suppression, there are no measurements of the non-singlet structure 
function for $x>0.9$ \cite{DISdata,Roberts:1992pe,Gehrmann:1999xn}. The 
experiments that probed the highest $x$-values were fixed-target experiments 
in the 1970s and 1980s at SLAC \cite{Whitlow:1991uw} and the BCDMS experiment 
at CERN \cite{Benvenuti:1989gs}. Newer experiments only cover $x\leq 0.65$. As 
a consequence, the threshold resummation in DIS is currently of limited 
phenomenological importance. However, the resummation is relevant for $W$ or 
Higgs production at hadron colliders, which can also be analyzed with the 
methods developed here.

\subsection{Parton evolution near the endpoint}
\label{subsec:Altarelli} 

The easiest way to derive the evolution equation for the parton distribution 
function in the limit $x\to 1$ is to use that the factorized expression 
(\ref{fact}) for the structure function $F_2$ must be independent of the 
arbitrary renormalization scale $\mu$, and to combine this information with 
the known scale dependences of the hard and jet functions, given in 
(\ref{gammaV}) and (\ref{Jrge}). This yields
\begin{eqnarray}\label{phievol}
   \frac{d}{d\ln\mu}\,\phi_q^{\rm ns}(\xi,\mu) 
   &=& 2\gamma^\phi(\alpha_s)\,\phi_q^{\rm ns}(\xi,\mu) 
    + 2\Gamma_{\rm cusp}(\alpha_s) \int_\xi^1\!d\xi'\,
    \frac{\phi_q^{\rm ns}(\xi',\mu)}{[\xi'-\xi]_*} \nonumber\\
   &=& \int_\xi^1\!\frac{dz}{z}\,P_{q\leftarrow q}^{({\rm endpt})}(z)\,
    \phi_q^{\rm ns}\Big( \frac{\xi}{z},\mu \Big) \,,
\end{eqnarray}
where
\begin{equation}\label{Pqq}
   P_{q\leftarrow q}^{({\rm endpt})}(z)
   = \frac{2\Gamma_{\rm cusp}(\alpha_s)}{(1-z)_+}
   + 2\gamma^\phi(\alpha_s)\,\delta(1-z)
\end{equation}
is the $z\to 1$ limit of the Altarelli-Parisi splitting function 
$P_{q\leftarrow q}(z)$, which is known from direct calculation at three-loop 
order \cite{Moch:2004pa}. The asymptotic form of the splitting function near 
the endpoint given above holds to all orders in perturbation theory, up to 
corrections of order $(1-z)$. Recall that the anomalous dimension 
$\gamma^\phi$ was defined as the difference of the anomalous dimensions 
$\gamma^J$ and $\gamma^V$ of the jet function and SCET current, see 
(\ref{gammaphi}). Relation (\ref{Pqq}) thus provides a check of our two-loop 
results for these anomalous dimensions, and it furthermore allows us to deduce 
the value of the three-loop coefficient $\gamma_2^J$ given in relation 
(\ref{gammaJ}) of the Appendix.

The exact solution to the evolution equation (\ref{phievol}) can be found in 
analogy with (\ref{jsol}). It reads
\begin{equation}
   \phi_q^{\rm ns}(\xi,\mu_f)
   = \exp\left[ 2a_{\gamma^\phi}(\mu_f,\mu_0) \right]
   \frac{e^{-\gamma_E\sigma}}{\Gamma(\sigma)}\,
   \int_\xi^1\!d\xi'\,
   \frac{\phi_q^{\rm ns}(\xi',\mu_0)}{(\xi'-\xi)^{1-\sigma}} \,,
\end{equation}
where this time $\sigma=2a_\Gamma(\mu_f,\mu_0)$, and $\mu_0$ denotes the scale 
at which the initial condition for $\phi_q^{\rm ns}$ is given. For the 
hadronic parameters $\cal N$ and $b$ governing the asymptotic behavior of the 
parton distribution function in (\ref{asymp}), this relation implies
\begin{eqnarray}
   b(\mu_f) &=& b(\mu_0) + 2a_\Gamma(\mu_f,\mu_0) \,, \nonumber\\
   {\cal N}(\mu_f) &=& {\cal N}(\mu_0)\,
    \exp\left[ 2a_{\gamma^\phi}(\mu_f,\mu_0) \right]
    \frac{e^{\gamma_E\,b(\mu_0)}\,\Gamma(1+b(\mu_0))}%
         {e^{\gamma_E\,b(\mu_f)}\,\Gamma(1+b(\mu_f))} \,.
\end{eqnarray}
These evolution equations ensure that the $\mu_f$ dependence on the two sides 
of relation (\ref{beauty}) is indeed the same. The first result is 
particularly simple and interesting. Since $a_\Gamma(\mu_f,\mu_0)>0$ for 
$\mu_f>\mu_0$, it follows that the coefficient $b(\mu)$ increases with $\mu$, 
a fact incompatible with the naive counting rule result $b=3$ 
\cite{Gunion:1973nm,Blankenbecler:1974tm}. In other words, such a counting 
rule could possibly hold only at a specific renormalization point.

\section{Connection with the standard approach}
\label{sec:connection}

The conventional approach to threshold resummation in DIS proceeds via moment 
space and inverse Mellin transformations \cite{Sterman:1986aj,Catani:1989ne}. 
The purpose of this section is twofold; first, to show that our momentum-space 
resummation is formally equivalent to the conventional resummation 
order by order in perturbation theory, and second, to point out and quantify 
the theoretical and numerical differences that appear in applications to 
physical quantities such as the DIS structure function.

We first recall some details of the conventional approach to threshold 
resummation in moment space. By taking moments of the structure function 
$F_2^{\rm ns}$, convolution integrals such as (\ref{simplefact}) or 
(\ref{fact}) can be brought into product form. The traditional way of writing
the result is 
\begin{equation}\label{F2Ndef}
   F_{2,N}^{\rm ns}(Q^2)
   = \int_0^1\!dx\,x^{N-1} F_2^{\rm ns}(x,Q^2)
   = C_N(Q^2,\mu_f)\,\sum_q\,e_q^2\,\phi_{q,N+1}^{\rm ns}(\mu_f) \,.
\end{equation}
where the moments of $C(Q^2,z,\mu)$ and $\phi_q^{\rm ns}(\xi,\mu)$ are defined 
in analogy to those of $F_2^{\rm ns}(x,Q^2)$. For large $N$, the function 
$C_N$ is then split into two pieces according to 
\begin{equation}\label{CN}
   C_N(Q^2,\mu_f) = g_0(Q^2,\mu_f)\,\exp\left[ G_N(Q^2,\mu_f) \right]
   + {\cal O}\Big( \frac{1}{N} \Big) \,,
\end{equation}
where $g_0$ contains all $N$-independent contributions, while the function 
$G_N$ contains logarithms of the form $\ln^k N$. The limit $x\to 1$ in 
momentum space corresponds to the limit $N\to\infty$ in moment space, so this 
formula achieves the exponentiation of large threshold logarithms. The 
resummation exponent $G_N$ is written as   
\begin{equation}\label{GN}
   G_N(Q^2,\mu_f) = \int_0^1\!dz\,\frac{z^{N-1}-1}{1-z} \left[
   \int_{\mu_f^2}^{(1-z)Q^2}\!\frac{dk^2}{k^2}\,A_q(\alpha_s(k))
   + B_q\left( \alpha_s(Q\sqrt{1-z}) \right) \right] \, , 
\end{equation}
where the functions $A_q$ and $B_q$ are universal radiation factors determined 
by matching with results from fixed-order perturbation theory. 

We shall now derive an equation relating the objects $g_0$, $A_q$, and $B_q$ 
in (\ref{CN}) and (\ref{GN}) to the matching coefficients and anomalous 
dimensions defined in effective field theory. We begin by transforming the 
factorization formula (\ref{fact}) into Mellin space, obtaining the product 
form 
\begin{equation}\label{F2N}
   F_{2,N}^{\rm ns}(Q^2)
   = |C_V(Q^2,\mu_f)|^2\,J_N(Q^2,\mu_f)\,
   \sum_q\,e_q^2\,\phi_{q,N+1}^{\rm ns}(\mu_f) \,,
\end{equation}
which is valid up to corrections in $1/N$. The Mellin-transformed jet function 
is defined as 
\begin{equation}
   J_N(Q^2,\mu)
   = \int_0^{Q^2}\!dp^2 \left( 1 - \frac{p^2}{Q^2} \right)^{N-1}
    J(p^2,\mu) \,.
\end{equation}
It was shown in \cite{Becher:2006qw} that for large $N$ the jet-function 
moments $J_N$ are given by
\begin{equation}
   J_N(Q^2,\mu) = \widetilde j\Big( \ln\frac{Q^2}{\bar N\mu^2},\mu \Big)
    + {\cal O}\Big( \frac{1}{N} \Big) \,, \qquad
   \bar N\equiv e^{\gamma_E} N \,, \label{JlargeN}
\end{equation}
and hence obey the same evolution equation (\ref{jtildeevol}) as the 
associated jet function. Using this connection along with the results derived 
in Section~\ref{sec:resummation}, the resummed coefficient function $C_N$ in 
(\ref{F2Ndef}) can be written as 
\begin{eqnarray}
   C_N(Q^2,\mu_f)
   &=& |C_V(Q^2,\mu_h)|^2
    \left( \frac{Q^2}{\mu_h^2} \right)^{-2a_\Gamma(\mu_h,\mu_i)}
    \exp\left[ 4S(\mu_h,\mu_i) - 2a_{\gamma^V}(\mu_h,\mu_i) \right] 
    \nonumber\\
   &&\times \exp\left[ 2a_{\gamma^\phi}(\mu_i,\mu_f) 
    - 2\ln\bar N\,a_\Gamma(\mu_i,\mu_f) \right]
    \widetilde j\Big( \ln\frac{Q^2}{\bar N\mu_i^2},\mu_i \Big)
    + {\cal O}\Big( \frac{1}{N} \Big) \,.
\end{eqnarray}

We now adopt the ``natural'' scale choices $\mu_h=Q$ and 
$\mu_i=Q/\sqrt{\bar N}$, which are implicit in most treatments of threshold 
resummation in the literature. This allows us to compare with the standard 
expression (\ref{CN}), but as we will discuss at the end of this section, this 
scale choice becomes problematic when the expressions for the moments are 
transformed back to $x$-space. Next, we express the RG functions 
$S(\mu_1,\mu_2)$ and $a_n(\mu_1,\mu_2)$ defined in (\ref{dgl}) in  terms of 
integrals over the appropriate anomalous dimensions. After a straightforward 
calculation, this leads to
\begin{eqnarray}\label{us}
   g_0^{\rm SCET}(Q^2,\mu_f)
   &=& |C_V(Q^2,Q)|^2\,\,\widetilde j(0,Q)\,
    \exp\left[ \int_{\mu_f^2}^{Q^2}\!\frac{dk^2}{k^2}\,
    \gamma^\phi(\alpha_s(k)) \right] , \nonumber\\
   G_N^{\rm SCET}(Q^2,\mu_f) &=& \int_{Q^2/\bar N}^{Q^2}\!\frac{dk^2}{k^2}
    \left[ \ln\frac{k^2}{Q^2}\,\Gamma_{\rm cusp}(\alpha_s(k))
    - \gamma^J(\alpha_s(k)) - \frac{d\ln\widetilde j(0,k)}{d\ln k^2} \right] 
    \nonumber\\
   &&\mbox{}- \ln\bar N \int_{\mu_f^2}^{Q^2/\bar N}\!\frac{dk^2}{k^2}\,
    \Gamma_{\rm cusp}(\alpha_s(k)) \,,
\end{eqnarray}
where we have defined the split between the two terms such that the expression 
for $G_N$ obtained in the large-$N$ limit vanishes for $\bar N\to 1$.

Our next task is to bring the exponent $G_N$ from the standard result 
(\ref{GN}) into a form resembling the SCET result (\ref{us}). Since the 
running coupling $\alpha_s(k)$ depends on its argument logarithmically, a 
helpful identity is (for integer $k\ge 0$)
\begin{equation}
   \int_0^1\!dz\,\frac{z^{N-1}-1}{1-z}\,\ln^k(1-z)
   = \frac{1}{k+1}\,I_{k+1}\Big( \ln\frac{1}{\bar N} \Big)
   + {\cal O}\Big( \frac{1}{N} \Big) \,,
\end{equation}
where 
\begin{equation}
   I_n(x) = \partial_\epsilon^n \left[ e^{\epsilon(x+\gamma_E)}\,
   \Gamma(1+\epsilon) \right]_{\epsilon\to 0}
\end{equation}
are $n$-th order polynomials defined in \cite{Neubert:2005nt}. With the help 
of these relations we find that for large $N$
\begin{eqnarray}\label{Vermaseren}
   G_N(Q^2,\mu_f) &=& \int_{Q^2/\bar N}^{Q^2}\!\frac{dk^2}{k^2} \left[
    \ln\frac{k^2}{Q^2}\,A_q(\alpha_s(k)) - B_q(\alpha_s(k)) \right] 
    + \Delta G\Big( \frac{Q}{\sqrt{\bar N}} \Big)  \nonumber\\
   &&\mbox{}- \ln\bar N \int_{\mu_f^2}^{Q^2/\bar N}\!\frac{dk^2}{k^2}\,
    A_q(\alpha_s(k)) + {\cal O}\Big( \frac{1}{N} \Big) \,,
\end{eqnarray}
where
\begin{equation}
   \Delta G(\mu) = \sum_{k=1}^\infty\,\frac{I_{k+1}(0)}{(k+1)!} 
   \left[ A_q^{(k-1)}(\alpha_s(\mu)) + B_q^{(k)}(\alpha_s(\mu)) \right] ,
\end{equation}
and $A_q^{(n)}$, $B_q^{(n)}$ denote the $n$-th derivatives of $A_q$ and $B_q$ 
with respect to $\ln\mu^2$. The perturbative expansion of $\Delta G$ starts at 
order $\alpha_s$. Contrary to the SCET expression in (\ref{us}), the result 
(\ref{Vermaseren}) does not vanish for $\bar N\to 1$. The overall 
normalization of $G_N(Q^2,\mu_f)$ is a matter of convention, since it can be 
absorbed into $g_0(Q^2,\mu_f)$. Taking the difference in normalization into 
account, the two definitions underlying (\ref{us}) and (\ref{Vermaseren}) are 
connected by 
\begin{eqnarray}
   g_0(Q^2,\mu_f) &=& g_0^{\rm SCET}(Q^2,\mu_f)\,
    \exp\left[ - \Delta G(Q) \right] , \nonumber\\
   G_N(Q^2,\mu_f) &=& G_N^{\rm SCET}(Q^2,\mu_f) + \Delta G(Q) \,.
\end{eqnarray}
At the expense of a proliferation of $\gamma_E$ terms in the perturbative 
expressions, one can equally well normalize $G_1(Q^2,\mu_f)=0$. This 
normalization condition is adopted, e.g., in \cite{Moch:2005ba}. 

Equation  (\ref{Vermaseren}) is consistent with (\ref{us}) if we identify 
$A_q(\alpha_s)=\Gamma_{\rm cusp}(\alpha_s)$ with the cusp anomalous dimension, 
and furthermore require that
\begin{equation}
   B_q(\alpha_s(\mu)) + \frac{d\Delta G(\mu)}{d\ln\mu^2}
   = \gamma^J(\alpha_s(\mu)) + \frac{d\ln\widetilde j(0,\mu)}{d\ln\mu^2} \,.
\end{equation}
This formula can be rearranged to read
\begin{equation}\label{Bqrela}
   e^{\gamma_E\nabla}\,\Gamma(1+\nabla)\,B_q(\alpha_s)
   = \gamma^J(\alpha_s) + \nabla\,\ln\widetilde j(0,\mu)
    - \left[ e^{\gamma_E\nabla}\,\Gamma(\nabla) - \frac{1}{\nabla} \right]
    \Gamma_{\rm cusp}(\alpha_s) \,,
\end{equation}
where
\begin{equation}
   \nabla = \frac{d}{d\ln\mu^2}
   = \frac{\beta(\alpha_s)}{2}\,\frac{\partial}{\partial\alpha_s} \,,
\end{equation}
and the differential operators are defined by their Taylor expansions. 
Evaluating (\ref{Bqrela}) in perturbation theory we obtain\footnote{A relation 
similar to (\ref{Bqrela}) has been derived in \cite{Idilbi:2006dg}; however, 
there is a typo in the last equation in (75) of that paper, which is the 
analog of our relation between $B_{q,3}$ and $\gamma_2^J$.}
\begin{eqnarray}\label{Bqpert}
   B_{q,1} &=& \gamma_0^J \,, \nonumber\\
   B_{q,2} &=& \gamma_1^J - \beta_0 b_0^{(1)} \,, \nonumber\\
   B_{q,3} &=& \gamma_2^J - \beta_1 b_0^{(1)}
    - \beta_0 \left[ 2b_0^{(2)} - \left( b_0^{(1)} \right)^2
    + \frac{\pi^2}{6} \left( \gamma_0^J \right)^2
    - 2\zeta_3\Gamma_0\gamma_0^J + \frac{\pi^4}{360}\,\Gamma_0^2 \right] ,
\end{eqnarray}
where the one- and two-loop matching coefficients $b_0^{(1)}$ and $b_0^{(2)}$ 
have been calculated in \cite{Becher:2006qw}. Computing the first three 
$B_{q,n}$ coefficients using the three-loop result for the anomalous dimension 
$\gamma^J$ given in the Appendix, we find agreement with the expressions 
derived in \cite{Moch:2005ba}. Ref.~\cite{Manohar:2003vb} identified the 
function $B_q$ with the jet-function anomalous dimension $\gamma^J$, which is 
incorrect already at two-loop order. 

Equations~(\ref{Bqrela}) and (\ref{Bqpert}) provide the desired relations 
between the function $B_q$ of the standard approach and the field-theoretical 
objects defined in the effective theory. Obviously, the connection between the 
various objects is highly non-trivial. This explains, perhaps, why it has 
proven difficult in the past to translate between the standard formalism and 
the approach based on SCET. The deeper reason is that in the conventional 
approach the RG evolution equations of SCET are replaced by a different set of 
partial differential equations \cite{Sterman:1986aj,Catani:1989ne}, whose 
solution is equivalent to our solution but not structurally identical to it. 
In particular, there are theoretical and numerical differences between the 
moment-space and momentum-space resummation procedures. Some of these are 
explicit in the particular form of the resummation exponent (\ref{GN}) 
obtained in moment space, and some become apparent only when performing the 
inverse Mellin transform. We conclude this section by examining these 
differences in more detail.

A troublesome feature of the conventional moment-space approach is that the 
integrals over the coupling constant in the resummation exponent run over the 
region where $\alpha_s(\mu)$ is evaluated at very small values of $\mu$. To 
leading order, the coupling behaves as 
\begin{equation}
   \alpha_s(\mu) = \frac{4\pi}{\beta_0\ln(\mu^2/\Lambda_{\rm QCD}^2)}
\end{equation}
and becomes infinite at the scale $\mu=\Lambda_{\rm QCD}$. When the 
integration variable $z$ in (\ref{GN}) approaches~1, the resummation exponent 
becomes sensitive to this Landau-pole singularity in the running coupling. As 
a result the integral is ambiguous, since one can arbitrarily choose a 
prescription for dealing with the pole. We can estimate the magnitude of the 
ambiguity by taking the difference of the $z$-integral evaluated above or 
below the Landau pole in (\ref{GN}). The result is
\begin{equation}
   \Delta G_N = - \frac{2\pi i}{\beta_0} 
   \left( \Gamma_0 + \gamma_0^J \right) (N-1)\,\frac{\Lambda_{\rm QCD}^2}{Q^2} 
   + {\cal O}\left( \frac{N^2\Lambda_{\rm QCD}^4}{Q^4} \right) ,
\end{equation}
which is of the form of a power correction of order 
$\Lambda_{\rm QCD}^2/M_X^2$. Note that this ambiguity never appears in the 
momentum-space formulation, and should therefore be interpreted as an artifact 
of resummation in moment space. As stressed earlier, the Landau-pole ambiguity 
does not imply an infrared renormalon-pole ambiguity of the same strength 
$\Lambda_{\rm QCD}^2/M_X^2$. To show that $G_N$ is indeed affected by a 
corresponding renormalon pole, one needs to evaluate the exponent in the 
large-$\beta_0$ limit, a fixed-order truncation of this quantity is not 
sufficient \cite{Beneke:1995pq}. On general grounds, one expects anomalous 
dimensions to be free of infrared renormalons, so that the renormalon poles 
enter only through the associated jet function $\widetilde j(0,\mu)$ in 
(\ref{Bqrela}). In the effective theory, renormalons affect only the matching 
coefficients, $C_V$ and $\widetilde j$, and will always be commensurate with 
power-suppressed operators. RG evolution, on the other hand, is driven by 
anomalous dimensions which are free of renormalons.
 
%%%%%%%%%%%%%%%%%%%%%%%%%%%%%%%%%%%%%%%%%%%%%%%%%%%%%%%%%%%%%%%%%%%
\begin{figure}
\begin{center}
\begin{tabular}{lr}
\hspace{-0.05\textwidth}
\psfrag{x}[B]{$x$}
\psfrag{y}[]{}
\psfrag{Q}[]{$Q=5$\,GeV}
\psfrag{T}[]{$K(x,Q^2,\mu_f)$}
\psfrag{n}[]{$M_X<1$\,GeV}
\includegraphics[width=0.48\textwidth]{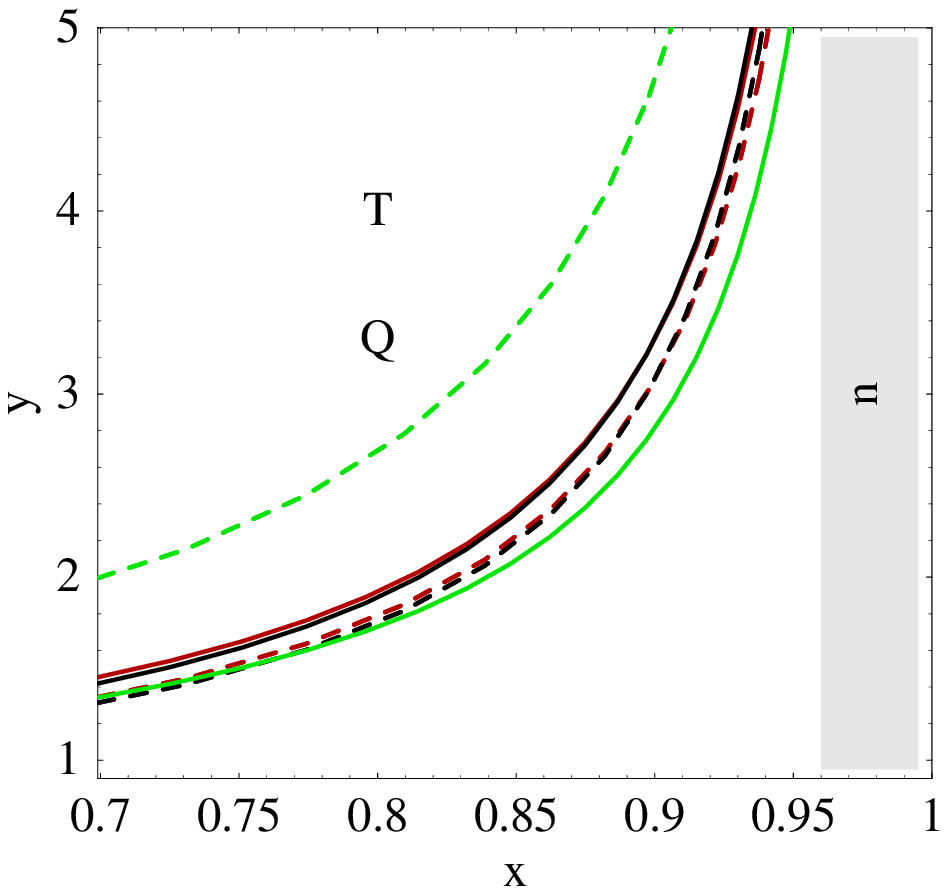} & 
\psfrag{x}[B]{$x$}
\psfrag{y}[]{}
\psfrag{Q}[]{$Q=30$\,GeV}
\psfrag{T}[]{$K(x,Q^2,\mu_f)$}
\includegraphics[width=0.48\textwidth]{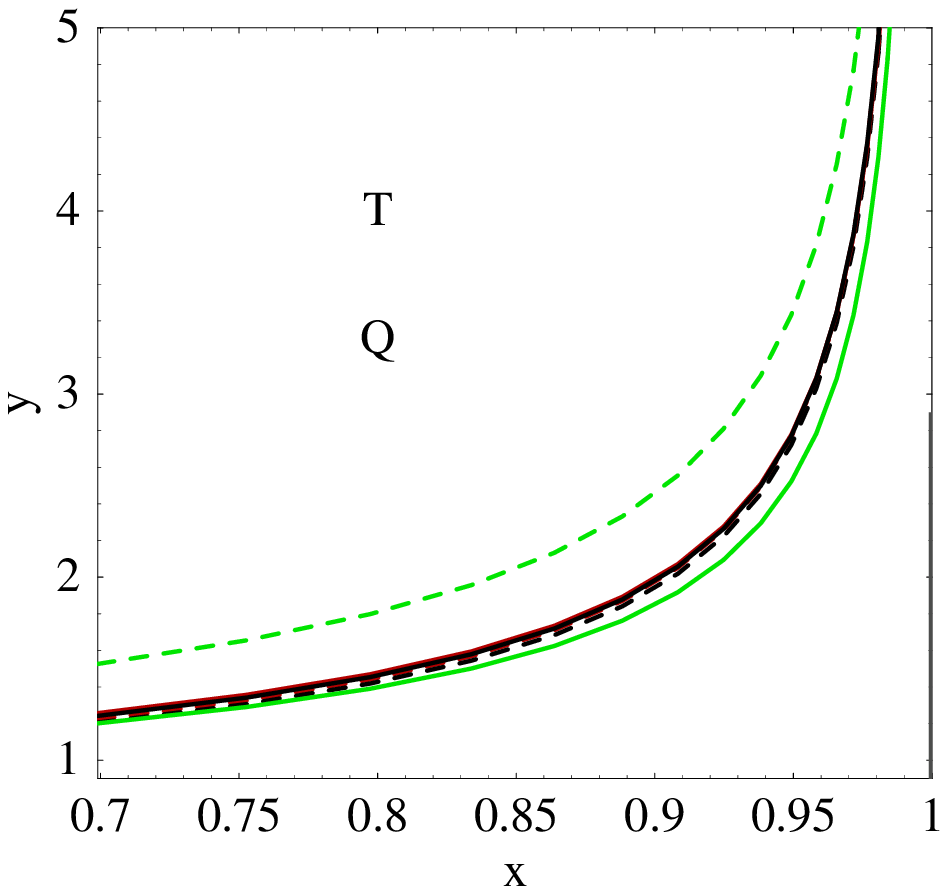}
\end{tabular}
\end{center}
\vspace{-0.5cm}
\caption{\label{comparisonMoment}
Comparison between Mellin-inverted moment space results (dashed) and results 
obtained in directly in $x$-space (solid). The green curves are the LO result, 
red NLO. The black lines are NNLO results and are visually indistinguishable 
from the NLO curves for $Q=30$\,GeV. We set $\mu_h=\mu_f=Q$, and $b(\mu_f)=4$. 
For the intermediate scale, we choose $\mu_i=M_X$ in momentum space and 
$\mu_i=Q/\sqrt{\bar N}$ in moment space.}
\end{figure}
%%%%%%%%%%%%%%%%%%%%%%%%%%%%%%%%%%%%%%%%%%%%%%%%%%%%%%%%%%%%%%%%%%%

Further differences become apparent when studying the inverse Mellin 
transformation needed to obtain the physical momentum-space results from the 
moment-space expressions. While our result (\ref{beauty}) obtained directly in 
$x$-space is completely analytical, the inverse Mellin transform can only be 
performed numerically, by evaluating the integral
\begin{equation}\label{invMell} 
   F_2^{\rm ns}(x,Q^2)
   = \frac{1}{2\pi i} \int_{c-i\infty}^{c+i\infty}\!dN\,x^{-N}
   C_N(Q^2,\mu_f)\,\phi_{q,N+1}^{\rm ns}(\mu_f) \,.
\end{equation}
The Mellin inversion is actually ambiguous, because the expression for $C_N$ 
has a Landau pole for large $N$. We deal with this pole by adopting the 
so-called minimal prescription \cite{Catani:1996yz}, which amounts to 
excluding the Landau pole from the integration contour by choosing the 
constant $c$ smaller than the value of $N$ at which the pole occurs. Even with 
this prescription, the numerical integral is not well behaved in the limit 
$x\to 1$, since the damping of the integrand becomes weaker and weaker as $x$ 
approaches the endpoint. In Figure~\ref{comparisonMoment} we compare the 
results for the $x$-space structure function obtained through numerical Mellin 
inversion with those obtained directly in  momentum space (\ref{beauty}). One 
source of numerical differences arises because the relation (\ref{JlargeN}) is 
only approximate,\footnote{The exact form of the RG equation obeyed by the 
jet-function moments can be found in \cite{Becher:2006qw}.}
so that the solution to the RG equation for $J_N(Q^2,\mu)$ receives 
corrections which are suppressed as $1/N$, while our momentum-space solution 
(\ref{sonice}) is exact. Another is that the default choice of the 
intermediate scale $\mu_i$ is different in the two approaches. The numerical 
differences are noticeable for smaller values of $Q$, but become negligible at 
$Q=30$\,GeV.

In the effective-theory result for the moments, the Landau pole in the inverse 
Mellin transformation can be avoided by performing the inversion to $x$-space 
with the appropriate scale choice for momentum space, 
$\mu_i\approx Q\sqrt{1-x}$, instead of $\mu_i= Q/\sqrt{\bar N}$. The freedom 
to choose the scales as appropriate for the quantity under consideration is an 
important advantage of our approach. The Landau-pole ambiguity in the Mellin 
inversion is not the only problem that arises from the fact that the scales 
cannot be varied in the standard resummation formalism. An additional 
difficulty was pointed out in \cite{Catani:1996yz}. To illustrate it, let us 
consider the structure function at the leading logarithmic level, even though 
this is not a consistent approximation in RG-improved perturbation theory. Our 
result (\ref{beauty}) then reduces to
\begin{equation}
   K(x,Q^2,\mu_f)
   = \exp\left[ 4S(\mu_h,\mu_i) + 2a_\Gamma(\mu_i,\mu_f)\ln(1-x)\right] ,
\end{equation}
where we have approximated $\frac{1-x}{x}\approx 1-x$. From (\ref{beauty}), we 
see that we have to choose $\mu_h\sim Q$ to make the double logarithms in the 
perturbative expansion of the hard matching coefficient $C_V(Q^2,\mu)$ small. 
Similarly, to avoid the appearance of large logarithms in the associated jet 
function $\widetilde j(\ln\frac{M_X^2}{\mu_i^2},\mu)$, the choice 
$\mu_i\sim M_X$ is mandatory. Let us now look at the structure function 
integrated over the endpoint region
\begin{equation}\label{endint}
   {\cal F}_2^{\rm ns}(x,Q^2) = \int_{1-x}^1\!dy\,F_2^{\rm ns}(y,Q^2) \,.
\end{equation}
In this case, the appropriate choice of the intermediate scale for integral 
${\cal F}_2^{\rm ns}(x,Q^2) $ is $\mu_i\sim Q\sqrt{1-x}$, as can be checked by 
explicitly performing the integral over (\ref{beauty}). If one instead chooses 
the scale $\mu_i$ to avoid logarithms on the level of the integrand, then the 
integral (\ref{endint}) becomes singular. To see the problem, we set 
$\mu_f=\mu_h=Q$, $\mu_i\approx Q\sqrt{1-y}$ and, for illustration purposes, 
approximate the Sudakov factor by expanding it to leading order around fixed 
coupling $\alpha_s(Q)$, as was done in \cite{Catani:1996yz}. The integral 
(\ref{endint}) then becomes
\begin{eqnarray}\label{div}
   {\cal F}_2^{\rm ns}(x,Q^2)
   &=& \int_{1-x}^1 dy\, \sum_q e_q^2\,y\,\phi_q^{\rm ns}(y,Q)\,
    \exp\left[ -a\ln^2\frac{\mu_i^2}{\mu_h^2}
    + 2a\ln\frac{\mu_i^2}{\mu_f^2}\,\ln(1-y) \right] \nonumber\\
   &=& \int_{1-x}^1 dy\,\sum_q e_q^2\,y\,\phi_q^{\rm ns}(y,\mu_f)\,
    \exp\left[a \ln^2(1-y)\right] ,
\end{eqnarray}
with $a=\Gamma_0 \frac{\alpha_s(Q)}{8\pi}$. Because the exponential factor 
grows faster than any power as $y\to 1$, this integral diverges. Its expansion 
in $a$ is an asymptotic series with factorially growing terms. As was shown in 
\cite{Catani:1996yz} the ambiguity associated with the non-integrable 
singularity for $y\to 1$ is of order 
\begin{equation}
   e^{-1/4a}\sim \Big(\frac{\Lambda_{\rm QCD}}{Q}\Big)^\frac{\beta_0}{4C_F}
   \approx \Big(\frac{\Lambda_{\rm QCD}}{Q}\Big)^{1.4}
\end{equation}
for $n_f=5$. In \cite{Catani:1996yz} it was shown that the above divergence 
does not occur if the Sudakov resummation is performed in moment space and the 
inverse transformation is performed exactly, without dropping subleading 
logarithms $\ln(1-x)$. From this, the authors concluded that the appropriate 
place to perform resummations is moment space and that leading logarithmic 
resummations in $x$-space are problematic. Our analysis shows that it is 
simply a bad choice of scale that produces the problem of the spurious power 
correction: the usual moment-space formalism produces logarithms $\ln^2 N$ in 
the Sudakov exponent, which translates into $\ln^2(1-x)$ at leading 
logarithmic accuracy, which in turn causes the problem in (\ref{div}). 
However, the proper way to perform the calculation is to keep the matching 
scales arbitrary and choose them such that the final result of a given 
calculation does not contain large logarithms. This avoids the above problem 
as well as the occurrence of Landau-pole ambiguities in inverse Mellin 
transforms. 

We hope that the above discussion helps to overcome the misconception that 
Mellin moment space is the ``correct place'' to perform the threshold 
resummation, and that resummation in $x$-space leads to inconsistent results. 
Quite to the contrary, the final analytical formulae we obtain in momentum 
space are simpler than those derived in moment space, they are free of 
spurious, unphysical power ambiguities and, as Figure~\ref{comparisonMoment} 
shows, the perturbative expansion in $x$-space exhibits a better apparent 
convergence.

\section{Summary and conclusions}

We analyzed DIS in the threshold region $x\to 1$ using SCET. With a detailed 
analysis in the effective theory, we rederived the standard QCD factorization 
theorem \cite{Sterman:1986aj,Catani:1989ne,Korchemsky:1992xv} for the 
non-singlet structure function $F_2^{\rm ns}(x,Q^2)$. While this process had 
been investigated in the effective theory before 
\cite{Manohar:2003vb,Pecjak:2005uh,Chay:2005rz,Manohar:2005az,Idilbi:2006dg,%
Chen:2006vd}, we argued that previous studies were incomplete. Our analysis 
resolves the issues left open in these papers. We agree with 
\cite{Pecjak:2005uh} that in a diagrammatic analysis momentum modes with low 
virtuality appear. Their presence is a consequence of the fact that in the 
limit $x\to 1$ one parton carries nearly all the momentum of the nucleon and 
the characteristic scale associated with the target remnants is $m^2(1-x)$, at 
least in a perturbative analysis. However, here we have shown that these modes 
do {\em not\/} translate into non-perturbative $Q$ dependence in the parton 
distribution function. In \cite{Manohar:2003vb} it was argued that such 
momentum regions do not contribute to the effective-theory calculation, and in 
\cite{Manohar:2005az} that they would be screened away non-perturbatively. 
Since these target remnants are part of the endpoint parton distribution 
function, it is incorrect to exclude them from the effective-theory 
factorization analysis, and there is no need to invoke a mechanism which 
eliminates them. For the same reason, there are no extra ``soft'' 
contributions outside the parton distribution function, contrary to what was 
postulated in \cite{Chay:2005rz}. 

With the factorization theorem at hand, we then performed the resummation 
of large logarithms by solving the RG equations of the effective theory. Our 
result involves three scales: a hard matching scale $\mu_h$, a jet scale 
$\mu_i$, and the factorization scale $\mu_f$. By choosing the perturbative 
scales to satisfy $\mu_h\sim Q$ and $\mu_i\sim M_X$, we avoid the presence of 
large logarithms. This approach has several advantages compared to the 
standard resummation technique for DIS. It enables us to derive a simple 
analytic expression for the resummed structure function directly in $x$-space, 
thus circumventing the problems associated with moment-space resummation. We 
can also estimate the higher-order perturbative uncertainties by varying the 
matching scales. It is trivial to recover the fixed-order result by setting 
all scales equal, $\mu_h=\mu_i=\mu_f$. This makes it straightforward to 
combine our results with fixed-order calculations valid away from the 
threshold region. 

An advantage of the effective-theory approach is that the resummed results are 
free of Landau-pole ambiguities. In the standard approach, these appear twice: 
in the resummed exponent in moment space and also in the Mellin inversion back 
to momentum space. Since we perform the resummation in momentum space by 
integrating out the higher scales and using RG evolution to go to lower 
scales, our expressions do not involve the strong coupling constant evaluated 
at scales below the minimum of the intermediate matching scale and the 
factorization scale at which the parton distribution function is renormalized. 
Therefore, Landau-pole ambiguities do not arise at any finite order in 
perturbation theory. We showed that our results are formally equivalent to the 
standard ones order by order in perturbation theory. This allowed us to relate 
the radiation function $B_q$ to a combination of the anomalous dimension 
$\gamma^J$ of the jet function and effective-theory matching coefficients. The 
two objects $B_q$ and $\gamma^J$ are identical at leading order, but beyond 
this the relation is highly non-trivial. 

Since the parton distribution functions fall off very rapidly near $x\to 1$, 
it is experimentally challenging to measure structure functions at large $x$. 
For this reason the amount of available experimental information near 
threshold is very limited. However, because of its relative simplicity and 
since the perturbative quantities are known with high precision, the threshold 
resummation for DIS has provided us with an ideal setup to develop our 
formalism. In the future, we plan to use the same approach to perform 
resummations in other, phenomenologically more relevant situations.

\subsection*{Acknowledgments}

We are grateful to Geoffrey Bodwin, Stefano Catani, and Einan Gardi for useful 
discussions. The research of T.B.\ was supported by the Department of Energy 
under Grant DE-AC02-76CH03000. The research of M.N.\ was supported by the 
National Science Foundation under Grant PHY-0355005. The work of B.P.\ was 
supported by the DFG Sonderforschungsbereich SFB/TR09 ``Computational 
Theoretical Particle Physics''. Fermilab is operated by Universities Research 
Association Inc., under contract with the U.S.\ Department of Energy.

\newpage
\section*{Appendix}

The exact solutions (\ref{RGEsols}) to the RG equations (\ref{dgl}) can be 
evaluated by expanding the anomalous dimensions and the QCD $\beta$-function 
as perturbative series in the strong coupling. We work consistently at NNLO in 
RG-improved perturbation theory, keeping terms through order $\alpha_s^2$ in 
the final expressions for the Sudakov exponent $S$ and the functions 
$a_\Gamma$, $a_{\gamma^V}$, and $a_{\gamma^J}$. We define the expansion 
coefficients as
\begin{eqnarray}
   \Gamma_{\rm cusp}(\alpha_s) &=& \Gamma_0\,\frac{\alpha_s}{4\pi}
    + \Gamma_1 \left( \frac{\alpha_s}{4\pi} \right)^2
    + \Gamma_2 \left( \frac{\alpha_s}{4\pi} \right)^3 
    + \Gamma_3 \left( \frac{\alpha_s}{4\pi} \right)^4 + \dots \,,
    \nonumber\\
   \beta(\alpha_s) &=& -2\alpha_s \left[ \beta_0\,\frac{\alpha_s}{4\pi}
    + \beta_1 \left( \frac{\alpha_s}{4\pi} \right)^2
    + \beta_2 \left( \frac{\alpha_s}{4\pi} \right)^3
    + \beta_3 \left( \frac{\alpha_s}{4\pi} \right)^4 + \dots \right] ,
\end{eqnarray}
and similarly for the other anomalous dimensions. In terms of these 
quantities, the function $a_\Gamma$ is given by
\begin{eqnarray}\label{asol}
   a_\Gamma(\nu,\mu)
   &=& \frac{\Gamma_0}{2\beta_0}\,\Bigg\{
    \ln\frac{\alpha_s(\mu)}{\alpha_s(\nu)}
    + \left( \frac{\Gamma_1}{\Gamma_0} - \frac{\beta_1}{\beta_0} \right)
    \frac{\alpha_s(\mu) - \alpha_s(\nu)}{4\pi} \nonumber\\ 
   &&\mbox{}+ \left[ \frac{\Gamma_2}{\Gamma_0} - \frac{\beta_2}{\beta_0}
    - \frac{\beta_1}{\beta_0}
    \left( \frac{\Gamma_1}{\Gamma_0} - \frac{\beta_1}{\beta_0} \right) \right]
    \frac{\alpha_s^2(\mu) - \alpha_s^2(\nu)}{32\pi^2} + \dots \Bigg\} \,.
\end{eqnarray}
The result for the Sudakov factor $S$ is more complicated. We obtain
\begin{eqnarray}
   S(\nu,\mu) &=& \frac{\Gamma_0}{4\beta_0^2}\,\Bigg\{
    \frac{4\pi}{\alpha_s(\nu)} \left( 1 - \frac{1}{r} - \ln r \right)
    + \left( \frac{\Gamma_1}{\Gamma_0} - \frac{\beta_1}{\beta_0}
    \right) (1-r+\ln r) + \frac{\beta_1}{2\beta_0} \ln^2 r \nonumber\\
   &&\mbox{}+ \frac{\alpha_s(\nu)}{4\pi} \Bigg[ 
    \left( \frac{\beta_1\Gamma_1}{\beta_0\Gamma_0} - \frac{\beta_2}{\beta_0} 
    \right) (1-r+r\ln r)
    + \left( \frac{\beta_1^2}{\beta_0^2} - \frac{\beta_2}{\beta_0} \right)
    (1-r)\ln r \nonumber\\
   &&\hspace{1.0cm}
    \mbox{}- \left( \frac{\beta_1^2}{\beta_0^2} - \frac{\beta_2}{\beta_0}
    - \frac{\beta_1\Gamma_1}{\beta_0\Gamma_0} + \frac{\Gamma_2}{\Gamma_0}
    \right) \frac{(1-r)^2}{2} \Bigg] \nonumber\\
   &&\mbox{}+ \left( \frac{\alpha_s(\nu)}{4\pi} \right)^2 \Bigg[
    \left( \frac{\beta_1\beta_2}{\beta_0^2} - \frac{\beta_1^3}{2\beta_0^3}
    - \frac{\beta_3}{2\beta_0} + \frac{\beta_1}{\beta_0}
    \left( \frac{\Gamma_2}{\Gamma_0} - \frac{\beta_2}{\beta_0}
    + \frac{\beta_1^2}{\beta_0^2} - \frac{\beta_1\Gamma_1}{\beta_0\Gamma_0}
    \right) \frac{r^2}{2} \right) \ln r \nonumber\\
   &&\hspace{1.0cm}
    \mbox{}+ \left( \frac{\Gamma_3}{\Gamma_0} - \frac{\beta_3}{\beta_0}
    + \frac{2\beta_1\beta_2}{\beta_0^2} + \frac{\beta_1^2}{\beta_0^2}
    \left( \frac{\Gamma_1}{\Gamma_0} - \frac{\beta_1}{\beta_0} \right)
    - \frac{\beta_2\Gamma_1}{\beta_0\Gamma_0}
    - \frac{\beta_1\Gamma_2}{\beta_0\Gamma_0} \right) \frac{(1-r)^3}{3}
    \nonumber\\
   &&\hspace{1.0cm}
    \mbox{}+ \left( \frac{3\beta_3}{4\beta_0} - \frac{\Gamma_3}{2\Gamma_0}
    + \frac{\beta_1^3}{\beta_0^3}
    - \frac{3\beta_1^2\Gamma_1}{4\beta_0^2\Gamma_0}
    + \frac{\beta_2\Gamma_1}{\beta_0\Gamma_0}
    + \frac{\beta_1\Gamma_2}{4\beta_0\Gamma_0}
    - \frac{7\beta_1\beta_2}{4\beta_0^2} \right) (1-r)^2 \nonumber\\
   &&\hspace{1.0cm}
    \mbox{}+ \left( \frac{\beta_1\beta_2}{\beta_0^2} - \frac{\beta_3}{\beta_0}
    - \frac{\beta_1^2\Gamma_1}{\beta_0^2\Gamma_0}
    + \frac{\beta_1\Gamma_2}{\beta_0\Gamma_0} \right) \frac{1-r}{2}
    \Bigg] + \dots \Bigg\} \,,
\end{eqnarray}
where $r=\alpha_s(\mu)/\alpha_s(\nu)$. Whereas the three-loop anomalous 
dimensions and $\beta$-function are required in (\ref{asol}), the expression 
for $S$ also involves the four-loop coefficients $\Gamma_3$ and $\beta_3$.

We now list expressions for the anomalous dimensions and the QCD 
$\beta$-function, quoting all results in the $\overline{{\rm MS}}$ 
renormalization scheme. For the convenience of the reader, we also give 
numerical results for $n_f=5$. The expansion of the cusp anomalous dimension
$\Gamma_{\rm cusp}$ to two-loop order was obtained some time ago 
\cite{Korchemskaya:1992je}, while recently the three-loop coefficient has been 
obtained in \cite{Moch:2004pa}. For the four-loop coefficient $\Gamma_3$, we 
use the Pad\'e approximants derived in \cite{Moch:2005ba}. The results are
\begin{eqnarray}
   \Gamma_0 &=& 4 C_F = \frac{16}{3} \,, \nonumber\\
   \Gamma_1 &=& 4 C_F \left[ \left( \frac{67}{9} - \frac{\pi^2}{3} \right)
    C_A - \frac{20}{9}\,T_F n_f \right] 
    \approx 36.8436 \,, \nonumber\\
   \Gamma_2 &=& 4 C_F \Bigg[ C_A^2 \left( \frac{245}{6} - \frac{134\pi^2}{27}
    + \frac{11\pi^4}{45} + \frac{22}{3}\,\zeta_3 \right) 
    + C_A T_F n_f  \left( - \frac{418}{27} + \frac{40\pi^2}{27}
    - \frac{56}{3}\,\zeta_3 \right) \nonumber\\
   &&\mbox{}+ C_F T_F n_f \left( - \frac{55}{3} + 16\zeta_3 \right) 
    - \frac{16}{27}\,T_F^2 n_f^2 \Bigg] 
    \approx 239.208 \,, \nonumber\\
   \Gamma_3 &\approx& 7849,~ 4313, ~ 1553 \quad \mbox{for} \quad
    n_f = 3,\,4,\,5 \,.
\end{eqnarray}
The anomalous dimension $\gamma^V$ can be determined up to three-loop order 
from the partial three-loop expression for the on-shell quark form factor in 
QCD, which has recently been obtained in \cite{Moch:2005id}. We find
\begin{eqnarray}
   \gamma_0^V &=& -6 C_F = - 8 \,, \nonumber\\
   \gamma_1^V &=& C_F^2 \left( -3 + 4\pi^2 - 48\zeta_3 \right)
    + C_F C_A \left( - \frac{961}{27} - \frac{11\pi^2}{3} + 52\zeta_3 \right)
    + C_F T_F n_f \left( \frac{260}{27} + \frac{4\pi^2}{3} \right) 
    \nonumber\\
    &\approx& 1.1419 \,, \nonumber\\
   \gamma_2^V &=& C_F^3 \left( -29 - 6\pi^2 - \frac{16\pi^4}{5}
    - 136\zeta_3 + \frac{32\pi^2}{3}\,\zeta_3 + 480\zeta_5 \right) \nonumber\\
   &&\mbox{}+ C_F^2 C_A \left( - \frac{151}{2} + \frac{410\pi^2}{9}
    + \frac{494\pi^4}{135} - \frac{1688}{3}\,\zeta_3
    - \frac{16\pi^2}{3}\,\zeta_3 - 240\zeta_5 \right) \nonumber\\
   &&\mbox{}+ C_F C_A^2 \left( - \frac{139345}{1458} - \frac{7163\pi^2}{243}
    - \frac{83\pi^4}{45} + \frac{7052}{9}\,\zeta_3
    - \frac{88\pi^2}{9}\,\zeta_3 - 272\zeta_5 \right) \nonumber\\
   &&\mbox{}+ C_F^2 T_F n_f \left( \frac{5906}{27} - \frac{52\pi^2}{9} 
    - \frac{56\pi^4}{27} + \frac{1024}{9}\,\zeta_3 \right) \nonumber\\
   &&\mbox{}+ C_F C_A T_F n_f \left( - \frac{34636}{729}
    + \frac{5188\pi^2}{243} + \frac{44\pi^4}{45} - \frac{3856}{27}\,\zeta_3 
    \right) \nonumber\\
   &&\mbox{}+ C_F T_F^2 n_f^2 \left( \frac{19336}{729} - \frac{80\pi^2}{27} 
    - \frac{64}{27}\,\zeta_3 \right)
    \approx -249.388 \,.
\end{eqnarray}
\newpage
\noindent
The results for the expansion coefficients of the jet-function anomalous 
dimension $\gamma^J$ are
\begin{eqnarray}\label{gammaJ}
   \gamma_0^J &=& -3 C_F = - 4 \,, \nonumber\\
   \gamma_1^J &=& C_F^2 \left( - \frac{3}{2} + 2\pi^2 - 24\zeta_3 \right) 
    + C_F C_A \left( - \frac{1769}{54} - \frac{11\pi^2}{9} + 40\zeta_3 \right)
    + C_F T_F n_f \left( \frac{242}{27} + \frac{4\pi^2}{9} \right)
    \nonumber\\ 
   &\approx & 38.6763 \,, \nonumber\\
   \gamma_2^J
   &=& C_F^3 \left( - \frac{29}{2} - 3\pi^2 - \frac{8\pi^4}{5} - 68\zeta_3 
    + \frac{16\pi^2}{3}\,\zeta_3 + 240\zeta_5 \right) \nonumber\\
   &&\mbox{}+ C_F^2 C_A \left( - \frac{151}{4} + \frac{205\pi^2}{9}
    + \frac{247\pi^4}{135} - \frac{844}{3}\,\zeta_3
    - \frac{8\pi^2}{3}\,\zeta_3 - 120\zeta_5 \right) \nonumber\\
   &&\mbox{}+ C_F C_A^2 \left( - \frac{412907}{2916} - \frac{419\pi^2}{243}
    - \frac{19\pi^4}{10} + \frac{5500}{9}\,\zeta_3
    - \frac{88\pi^2}{9}\,\zeta_3 - 232\zeta_5 \right) \nonumber\\
   &&\mbox{}+ C_F^2 T_F n_f \left( \frac{4664}{27} - \frac{32\pi^2}{9}
    - \frac{164\pi^4}{135} + \frac{208}{9}\,\zeta_3 \right) \nonumber\\
   &&\mbox{}+ C_F C_A T_F n_f \left( - \frac{5476}{729}
    + \frac{1180\pi^2}{243} + \frac{46\pi^4}{45}
    - \frac{2656}{27}\,\zeta_3 \right) \nonumber\\
   &&\mbox{}+ C_F T_F^2 n_f^2 \left( \frac{13828}{729} - \frac{80\pi^2}{81}
    - \frac{256}{27}\,\zeta_3 \right)
    \approx 204.816 \,.
\end{eqnarray}
Finally, the expansion coefficients for the QCD $\beta$-function to four-loop 
order are
\begin{eqnarray}
   \beta_0 &=& \frac{11}{3}\,C_A - \frac43\,T_F n_f 
    = \frac{23}{3} \,, \nonumber\\
   \beta_1 &=& \frac{34}{3}\,C_A^2 - \frac{20}{3}\,C_A T_F n_f
    - 4 C_F T_F n_f \approx 38.6667\,, \\
   \beta_2 &=& \frac{2857}{54}\,C_A^3 + \left( 2 C_F^2
    - \frac{205}{9}\,C_F C_A - \frac{1415}{27}\,C_A^2 \right) T_F n_f
    + \left( \frac{44}{9}\,C_F + \frac{158}{27}\,C_A \right) T_F^2 n_f^2
    \nonumber\\
   &\approx & 180.907 \,, \nonumber\\
   \beta_3 &=& \frac{149753}{6} + 3564\zeta_3
    - \left( \frac{1078361}{162} + \frac{6508}{27}\,\zeta_3 \right) n_f
    + \left( \frac{50065}{162} + \frac{6472}{81}\,\zeta_3 \right) n_f^2
    + \frac{1093}{729}\,n_f^3 \nonumber\\
    &\approx& 4826.16 \,, 
\end{eqnarray}
where the value of $\beta_3$ is taken from \cite{vanRitbergen:1997va} and 
corresponds to $N_c=3$ and $T_F=\frac12$.

\end{document}